\documentclass{emulateapj}
\usepackage{color}
\usepackage{amsmath}
\usepackage{natbib}
\shorttitle{Radio Telescope Gain and Polarization Properties}
\shortauthors{Du et al.}

\begin{document}

\title{Gain and Polarization Properties of a Large Radio Telescope from
Calculation and Measurement: The John A. Galt Telescope}

\author{X. Du\altaffilmark{1,2,3}, T. L. Landecker\altaffilmark{1}, T.
Robishaw\altaffilmark{1},  A. D. Gray\altaffilmark{1}, K. A.
Douglas\altaffilmark{1,4,5}, M. Wolleben\altaffilmark{1,6}}

\email{duxuanmax@gmail.com}

\altaffiltext{1}{National Research Council Canada, Dominion Radio Astrophysical Observatory, P.O. Box 248, Penticton, British Columbia, V2A 6J9, Canada}

\altaffiltext{2}{Department of Electrical Engineering, University of Victoria, Canada}

\altaffiltext{3}{School of Engineering, University of British Columbia, Kelowna, Canada}

\altaffiltext{4}{Department of Physics and Astronomy, University of Calgary, 2500 University Drive, Calgary, AB, T2N 1N4, Canada}

\altaffiltext{5}{Physics and Astronomy Department, Okanagan College, 1000 KLO Road, Kelowna, British Columbia, V1Y 4X8, Canada}

\altaffiltext{6}{Skaha Remote Sensing Ltd., 3165 Juniper Drive, Naramata, BC, V0H 1N0, Canada}

\begin{abstract}
Measurement of the brightness temperature of extended radio emission demands knowledge of the gain (or aperture efficiency) of the telescope and measurement of the polarized component of the emission requires correction for the conversion of unpolarized emission from sky and ground to apparently polarized signal. Radiation properties of the John A. Galt Telescope at the Dominion Radio Astrophysical Observatory were studied through analysis and measurement in order to provide absolute calibration of a survey of polarized emission from the entire northern sky from 1280 to 1750 MHz, and to understand the polarization performance of the telescope. Electromagnetic simulation packages CST and GRASP-10 were used to compute complete radiation patterns of the telescope in all Stokes parameters, and thereby to establish gain and aperture efficiency. Aperture efficiency was also evaluated using geometrical optics and ray tracing analysis and was measured based on the known flux density of Cyg A. Measured aperture efficiency varied smoothly with frequency between values of 0.49 and 0.54; GRASP-10 yielded values 6.5\% higher but with closely similar variation with frequency. Overall error across the frequency band is 3\%, but values at any two frequencies are relatively correct to $\sim$1\%. Dominant influences on aperture efficiency are the illumination taper of the feed radiation pattern and the shadowing of the reflector from the feed by the feed-support struts. A model of emission from the ground was developed based on measurements and on empirical data obtained from remote sensing of the Earth from satellite-borne telescopes. This model was convolved with the computed antenna response to estimate conversion of ground emission into spurious polarized signal. The computed spurious signal is comparable to measured values, but is not accurate enough to be used to correct observations. A simpler model, in which the ground is considered as an unpolarized emitter with a brightness temperature of $\sim$240 K, is shown to have useful accuracy when compared to measurements.

\end{abstract}


\section{Introduction}
\label{intro}

Full exploitation of a radio telescope requires detailed and accurate knowledge of its radiation properties. Modern tools of antenna engineering, electromagnetic simulators, offer the ability to analyze telescope performance. To what extent can they deliver a true picture of telescope characteristics? Can we use them to calculate telescope behavior to useful accuracy? In this paper we investigate these questions by analyzing the properties of a large radio telescope and comparing the computations with measurements. The telescope that we study is the John A. Galt Telescope at the Dominion Radio Astrophysical Observatory.

We focus on two problems particularly relevant to mapping the extended emission from the Milky Way, which is partially linearly polarized at decimeter wavelengths. First, we need to know the gain (or, equivalently, the aperture efficiency) of the telescope so that we can report our observations in units of brightness temperature. Second, we need to know the polarization behavior of the telescope so that we can correct observations for instrumental polarization.

The John A. Galt Telescope (we will refer to it as the Galt Telescope) is a 26 m axially symmetric paraboloidal reflector. The Galt Telescope has recently been used to map the polarized emission from the entire northern sky over the frequency range 1280 to 1750~MHz as part of the Global Magneto-Ionic Medium Survey (GMIMS -- \citealp{woll09}). The results reported here have been used in processing the data from that survey.

The challenge in astronomical polarimetry is the accurate measurement of a small polarized signal embedded in a (usually) much larger randomly polarized signal.{\footnote{A randomly polarized signal is one whose time average has no net polarization} Instrumental polarization has a deleterious effect on polarimetry because it converts an unpolarized signal into an apparently polarized one. Radiation from the ground is unavoidable in most reflector telescopes, and this emission is similarly converted into an unwanted polarized signal, usually a substantial one. We examine the properties of ground emission and the telescope response to it.

To attain our two goals we compute the total radiation pattern of the telescope, including its polarization response. We describe telescope response in terms of Stokes parameters, widely used in astronomy to characterize the polarization state of a signal \citep{tinb96,wils14} but less frequently used in antenna engineering. Stokes parameter $I$ is proportional to the total intensity of the signal, parameters $Q$ and $U$ together describe the state of linear polarization and parameter $V$ describes the state of circular polarization. For a linearly polarized signal the polarized intensity is ${\rm PI}=\sqrt{{Q^2}+{U^2}}$ and the polarization angle is ${\alpha}={0.5~{{\rm{tan}}^{-1}({U/Q})}}$. Positive $V$ corresponds to right-hand circular polarization (RHCP) and negative $V$ to left-hand circular polarization (LHCP) \citep{ieee79}.

The terminology of antenna engineering is rife with terms that were developed while considering the antenna as a transmitter, for example the use of {\it{feed}} to describe the antenna placed at the focus of a reflector, or {\it{spillover}} to describe radiation from the feed that enters the far field without encountering the reflector. We calculate the radiation properties of the Galt Telescope as a transmitter, and can confidently use the results to understand its behavior as a receiver because reciprocity informs us that the behavior of an antenna as a receiver is completely described by its properties as a transmitter. Our calculations assume that the antenna is transmitting a signal at a single frequency, but we apply our results to receiving the wideband noise signals of radio astronomy.

In describing radiation patterns we use the terms $E$ plane and $H$ plane. The $E$ plane is the plane which contains the axis of a linearly polarized feed (or antenna) and the electric vector of the excitation. The $H$ plane is orthogonal to the $E$ plane, and also contains the antenna axis.

\section{Calculating the Radiation Pattern}
\label{calc}

\subsection{Calculating the Radiation Properties of the Feed}
\label{feed}

The feed is based on the design of \citet{wohl72}, scaled to a center frequency of 1576~MHz. The radiation properties of the feed were calculated using the CST software package \citep{cst14}; representative radiation patterns are shown in Figure~\ref{primary}. At the nominal center frequency the $E$- and $H$-plane patterns have nearly equal width and closely match the patterns measured by \citet{wohl72} on a 2800 MHz version of the feed. At frequencies below the design frequency of 1576~MHz the $E$-plane pattern is narrower than the $H$-plane pattern, and above that frequency the reverse is true.

\begin{figure}[ht!]
   \centerline{\includegraphics[width=0.48\textwidth]{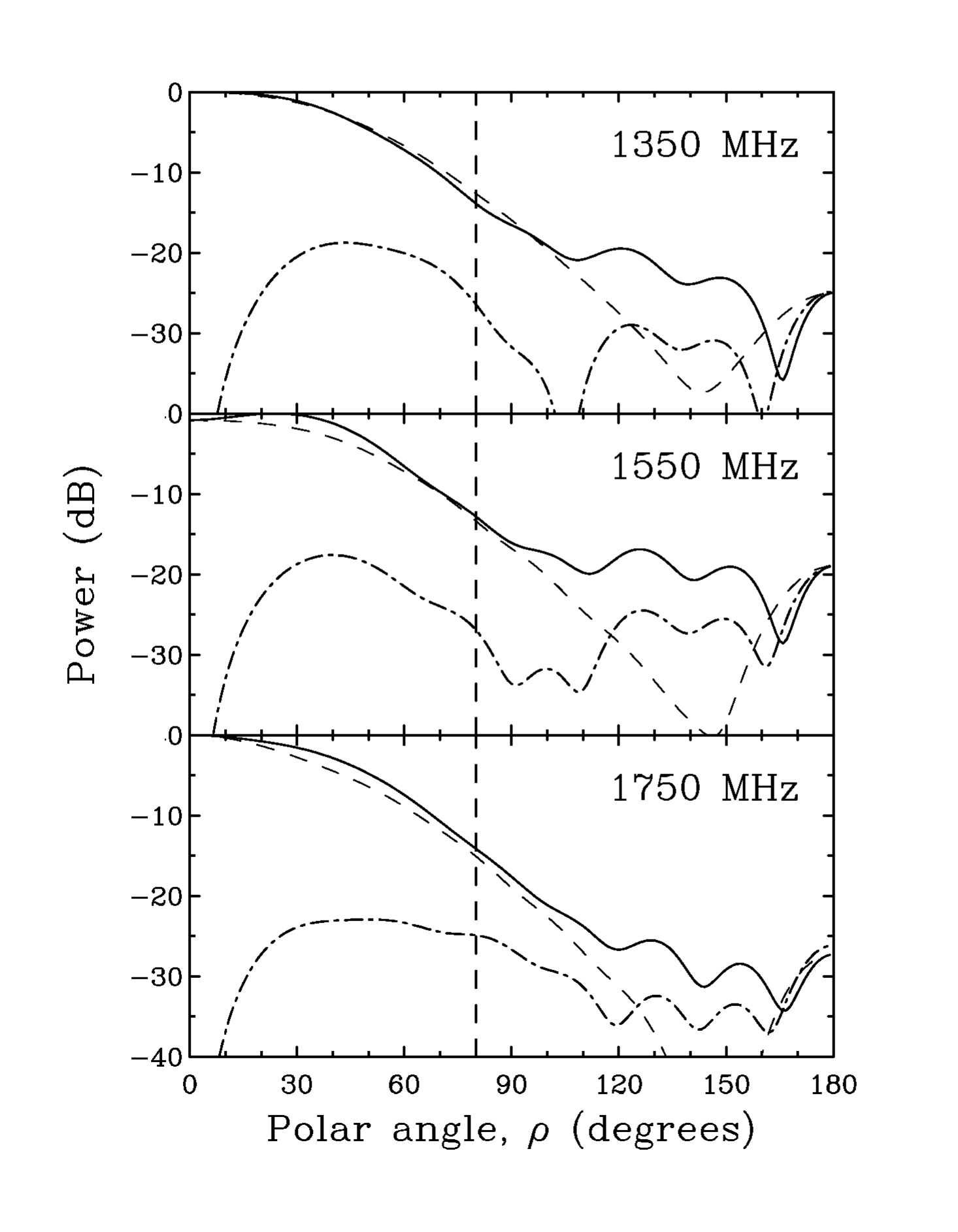}}
   \caption{Radiation patterns of the feed as a function of the polar angle $\rho$ at the indicated frequencies,
   calculated using CST. Solid curves: $E$-plane patterns. Dashed curves: $H$-plane
   patterns. Dot-dash curves: cross-polar patterns in the $45^{\circ}$ plane.
   The vertical lines mark the outer edge of the reflector at
   ${\rho}={80^{\circ}}$.}
   \label{primary}
\end{figure}

\subsection{Calculating the Radiation Properties of the Telescope}
\label{grasp}

Our computations used the software package GRASP-10, version 10.1.0 \citep{gras12}, which employs physical optics (PO) and the physical theory of diffraction (PTD) to calculate the far field radiated from the telescope. The calculation includes the effects of the feed, feed-support struts, and reflector, taking account of aperture blockage. CST provides the incident field from the feed, including its full polarization characteristics, as input to the calculation. GRASP-10 uses this input to calculate surface currents on the basis of PO, modified by PTD which models the current near the rim of the reflector. The radiation pattern of the telescope is then calculated using those currents. Therefore, GRASP-10 accurately models shadowing and the effects commonly attributed to diffraction.

The reflector has a focal-length-to-diameter ratio of 0.298, and subtends $160.1^{\circ}$ at the feed. The feed struts make an angle of $34^{\circ}$ with the telescope axis. Because of limitations of GRASP-10 the feed struts in the model do not contact the reflector surface or the feed (the feed is not a physical structure in GRASP-10, but merely a radiation source with known properties). Furthermore, the struts are modelled as metal cylinders, when in fact they are made of fiberglass, have a cigar shape, and two of them have metal-sheathed cables running along them.\footnote{After version 10.3.0 GRASP is able to model dielectric feed support struts.} We will discuss the impact of these approximations in Section~\ref{disc}.

Each GRASP-10 calculation follows this procedure: (a) scattered fields from the struts, induced by the spherical wave emanating from the feed, are calculated, (b) the reflected (plane) wave from the paraboloidal surface is calculated, taking into account both the spherical wave from the feed and the fields scattered from the struts, (c) the fields scattered by the struts are calculated, as the struts are illuminated by the plane wave from the reflector, and (d) fields scattered from the struts and the fields of the plane wave from the reflector are added. Note that the struts scatter radiation {\it{twice}}. The work was extended by including the effects of departures of the reflector surface from a perfect paraboloid (surface roughness), and of leakage of radiation through the mesh surface of the reflector. Computations were made at 50 MHz intervals from 1250 to 1750~MHz.

The output of GRASP-10 contained the real and imaginary components of the electric far field, $E$. The Stokes parameters were calculated as
\begin{subequations}\label{stokes}
\begin{align}
    I &= {[{\rm Re}(E_x)]^2} + {[{\rm Im}(E_x)]^2} + {[{\rm Re}(E_y)]^2} + {[{\rm Im}(E_y)]^2} \thinspace,\\
%\end{equation}
%\begin{equation}
    Q &= {[{\rm Re}(E_x)]^2} + {[{\rm Im}(E_x)]^2} - {[{\rm Re}(E_y)]^2} - {[{\rm Im}(E_y)]^2} \thinspace,\\
%\end{equation}
%\begin{equation}
    U &= 2\thinspace{\rm Re}(E_x){\thinspace}{\rm Re}(E_y) + 2\thinspace{\rm Im}(E_x){\thinspace}{\rm Im}(E_y) \thinspace,\\
%\end{equation}
%\begin{equation}
    V &= 2\thinspace {\rm Im}(E_x){\thinspace}{\rm Re}(E_y) - 2\thinspace{\rm Re}(E_x){\thinspace}{\rm Im}(E_y) \thinspace,
%\end{equation}
\end{align}
\end{subequations}
where $x$ and $y$ are the co- and cross-polar directions.

We need to calculate the response of the telescope to an incoming unpolarized signal. Since we treat the antenna as a transmitter, we need to simulate an antenna that transmits an unpolarized signal.{\footnote{An unpolarized transmitter is a theoretical concept: all real antennas are polarized, and therefore all transmitted signals are polarized}} Random polarization is simulated by rotating the feed to four positions in $45^{\circ}$ steps and averaging the resulting $I$, $Q$, $U$, and $V$ patterns: this technique was
developed by \citet{ng05}. Thus
\begin{equation}
{Q_{\rm random}}=( {{Q_{45^{\circ}}}+{Q_{90^{\circ}}}+{Q_{135^{\circ}}}+{Q_{180^{\circ}}}} )/4 \thinspace,
\end{equation}
and similarly for the other Stokes parameters.

\section{Results - Radiation Patterns}
\label{radpa}

\begin{figure*}[ht]
    \centerline{
    \includegraphics[bb = 195 150 598 596,width=0.27\textwidth,clip]
    {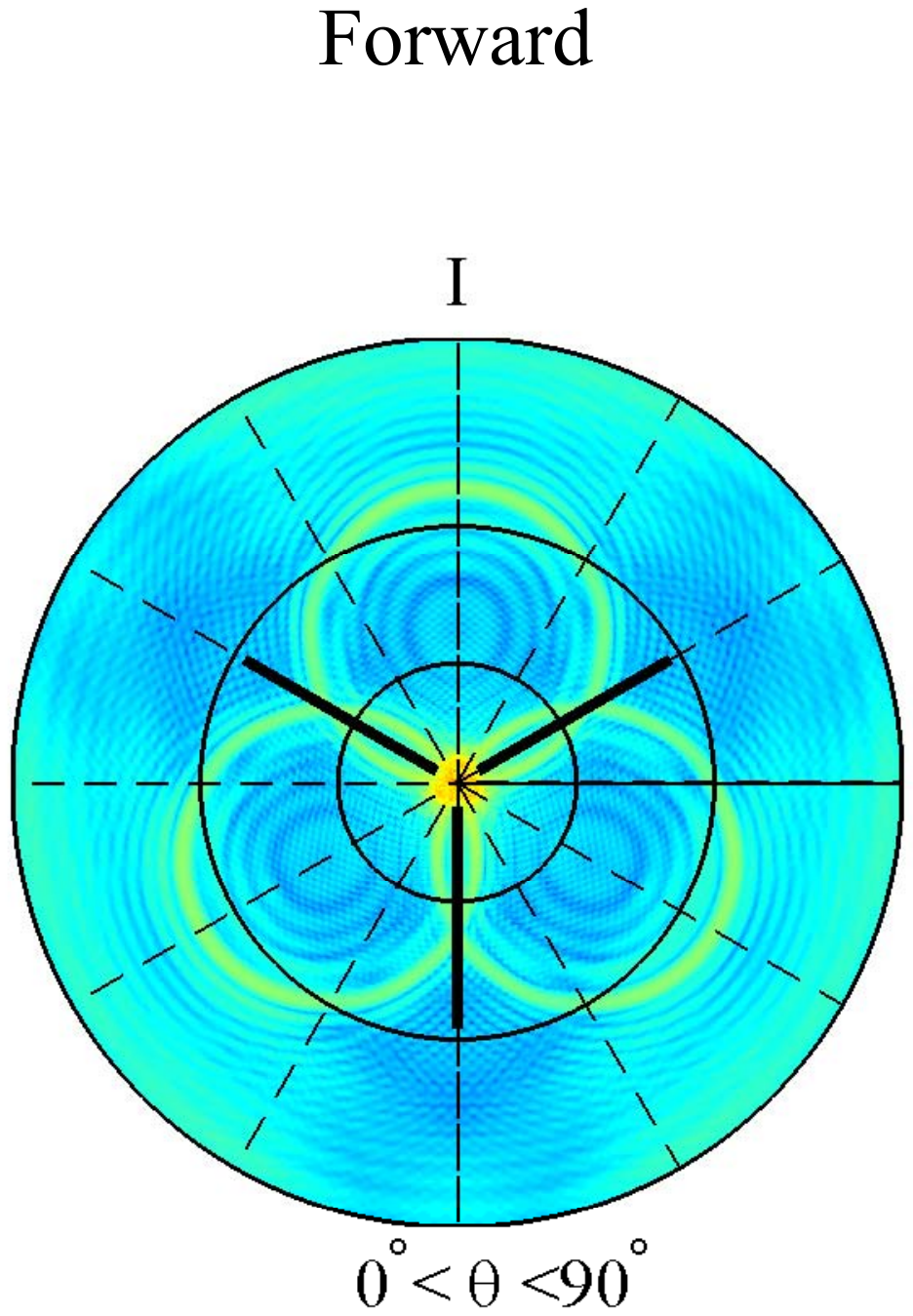}
    \includegraphics[bb = 195 150 598 596,width=0.27\textwidth,clip]
    {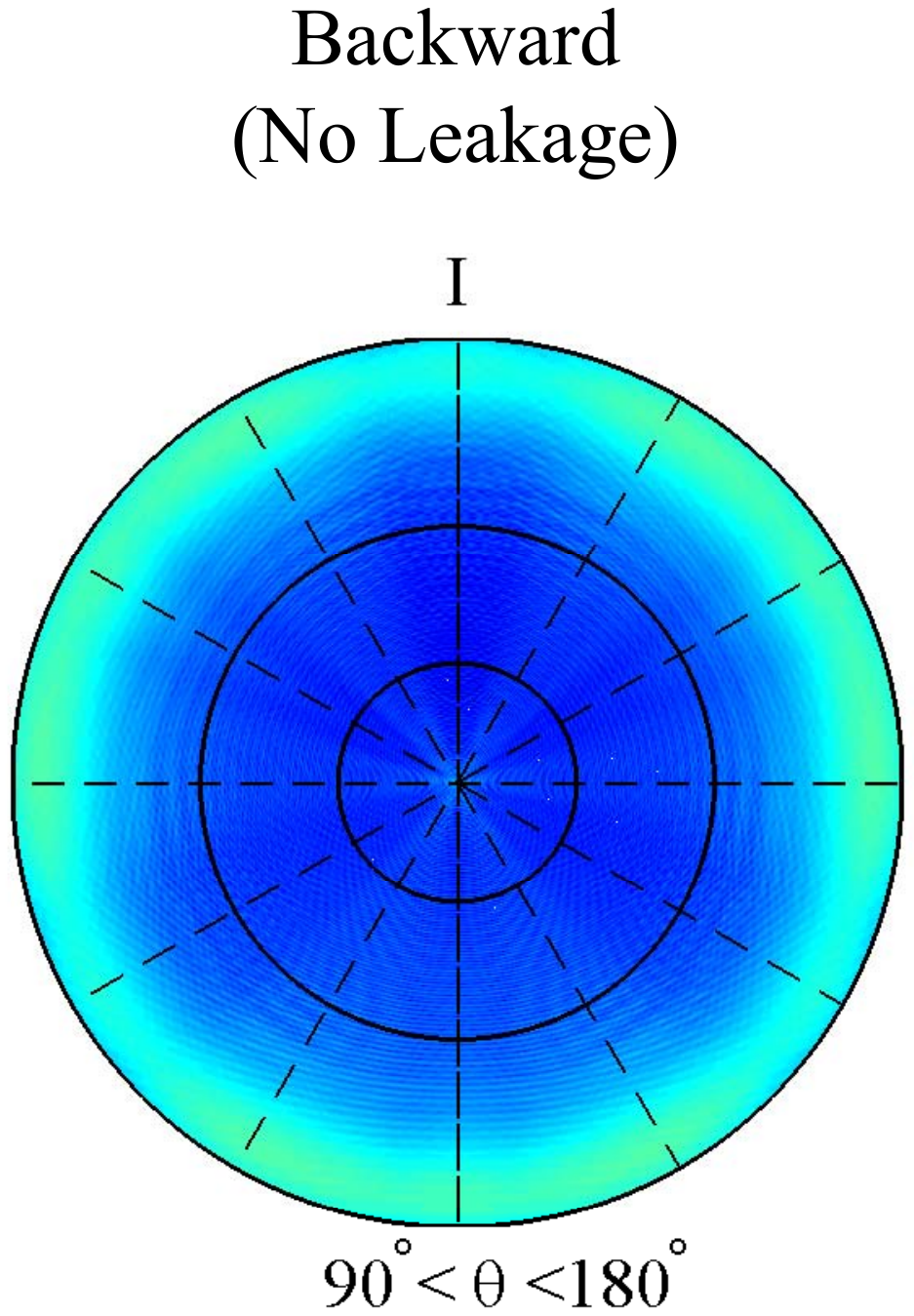}
    \includegraphics[bb = 195 150 598 596,width=0.27\textwidth,clip]
    {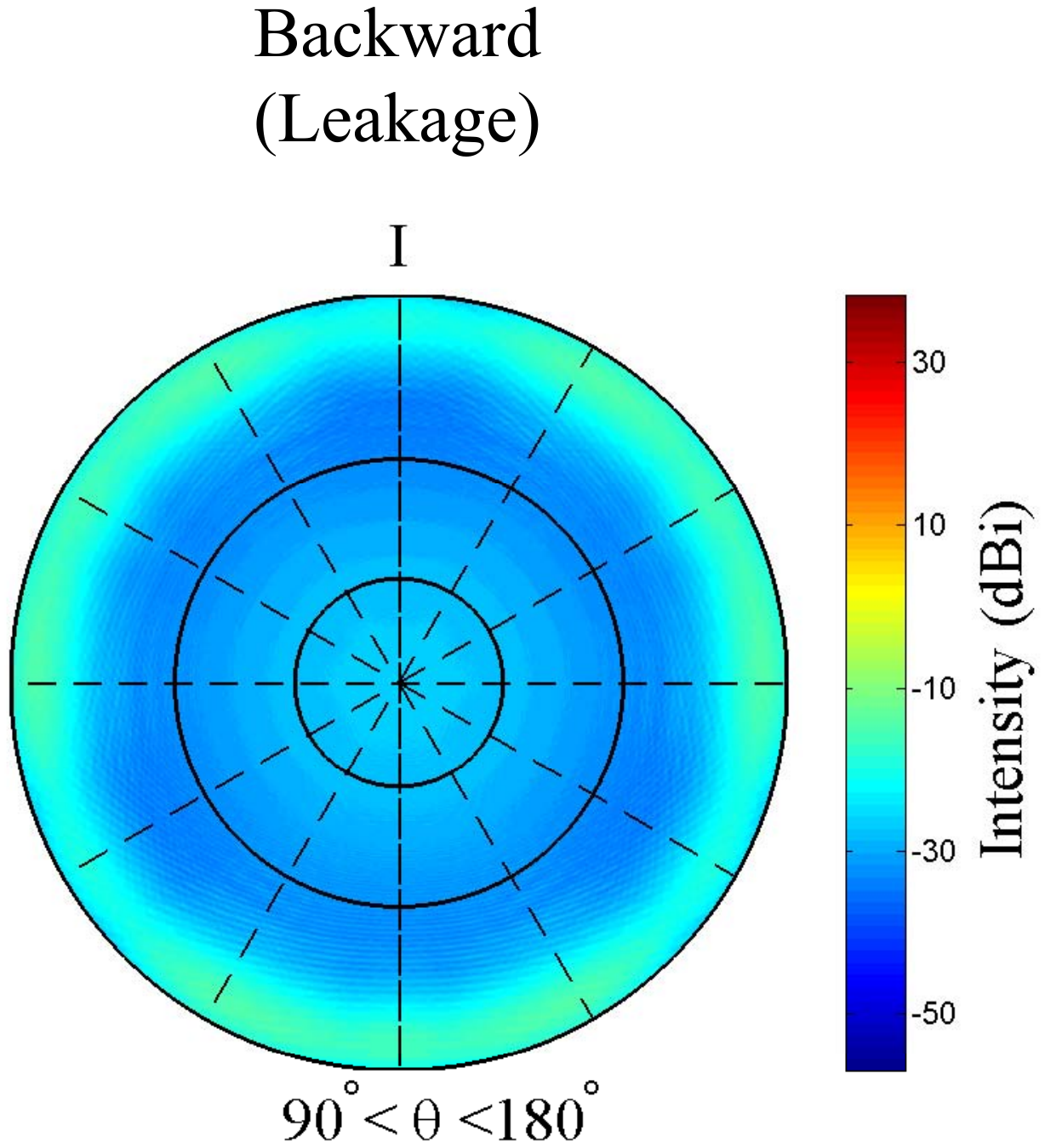}}
    \centerline{
    \includegraphics[bb = 195 150 598 470,width=0.27\textwidth,clip]
    {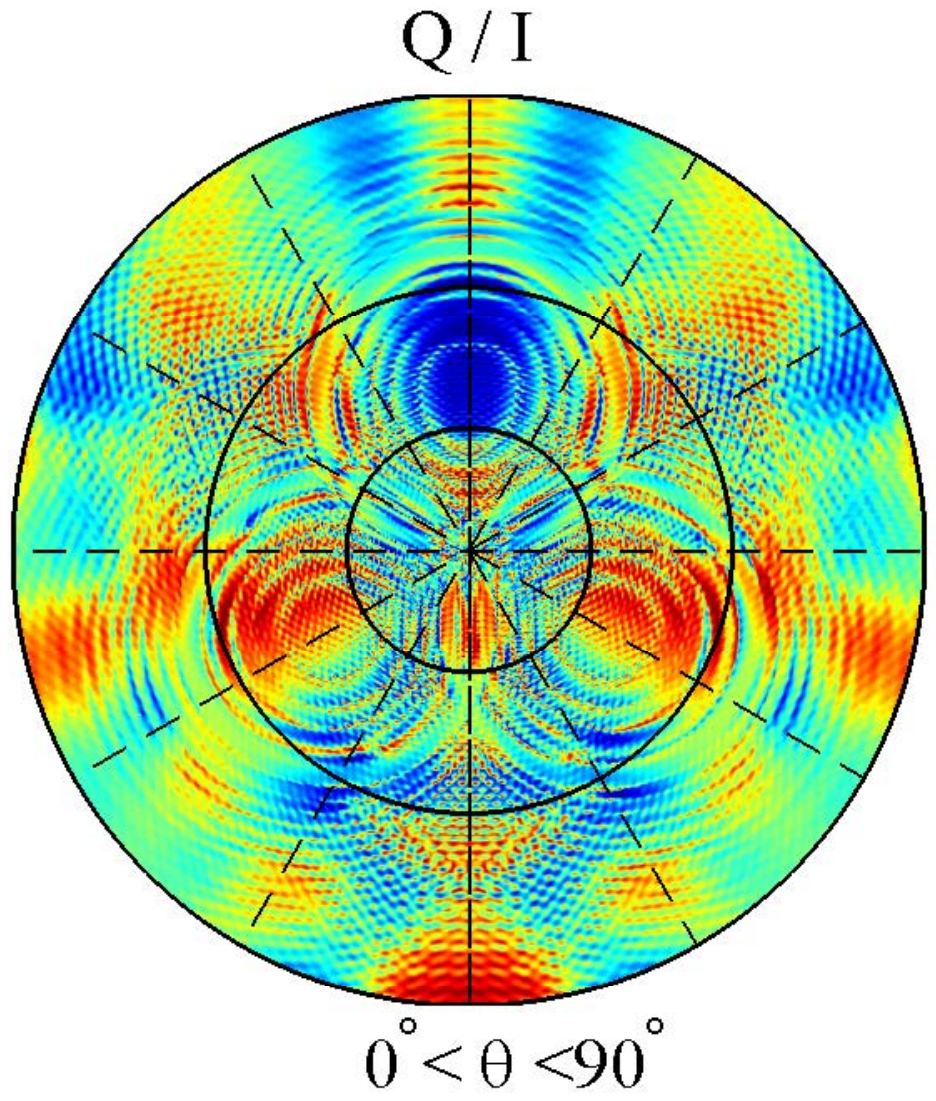}
    \includegraphics[bb = 195 150 598 470,width=0.27\textwidth,clip]
    {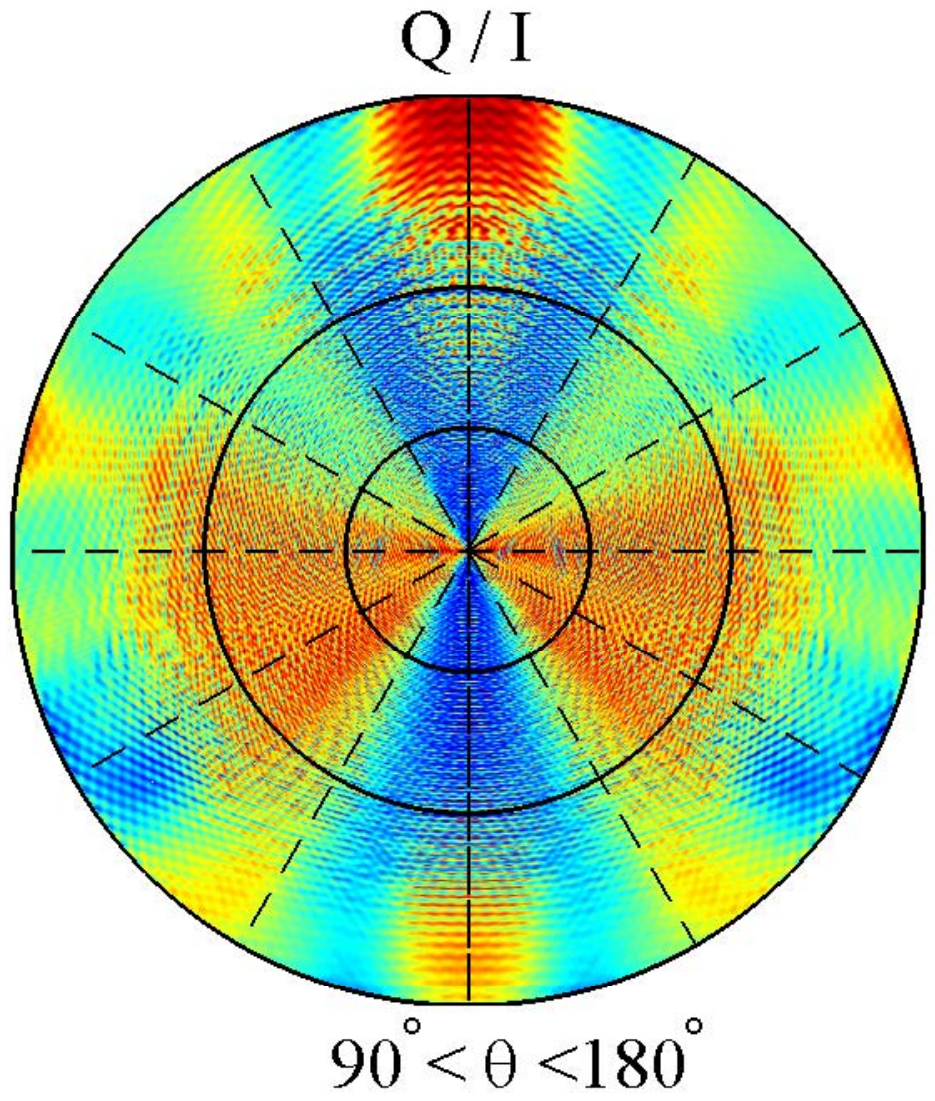}
    \includegraphics[bb = 195 150 598 470,width=0.27\textwidth,clip]
    {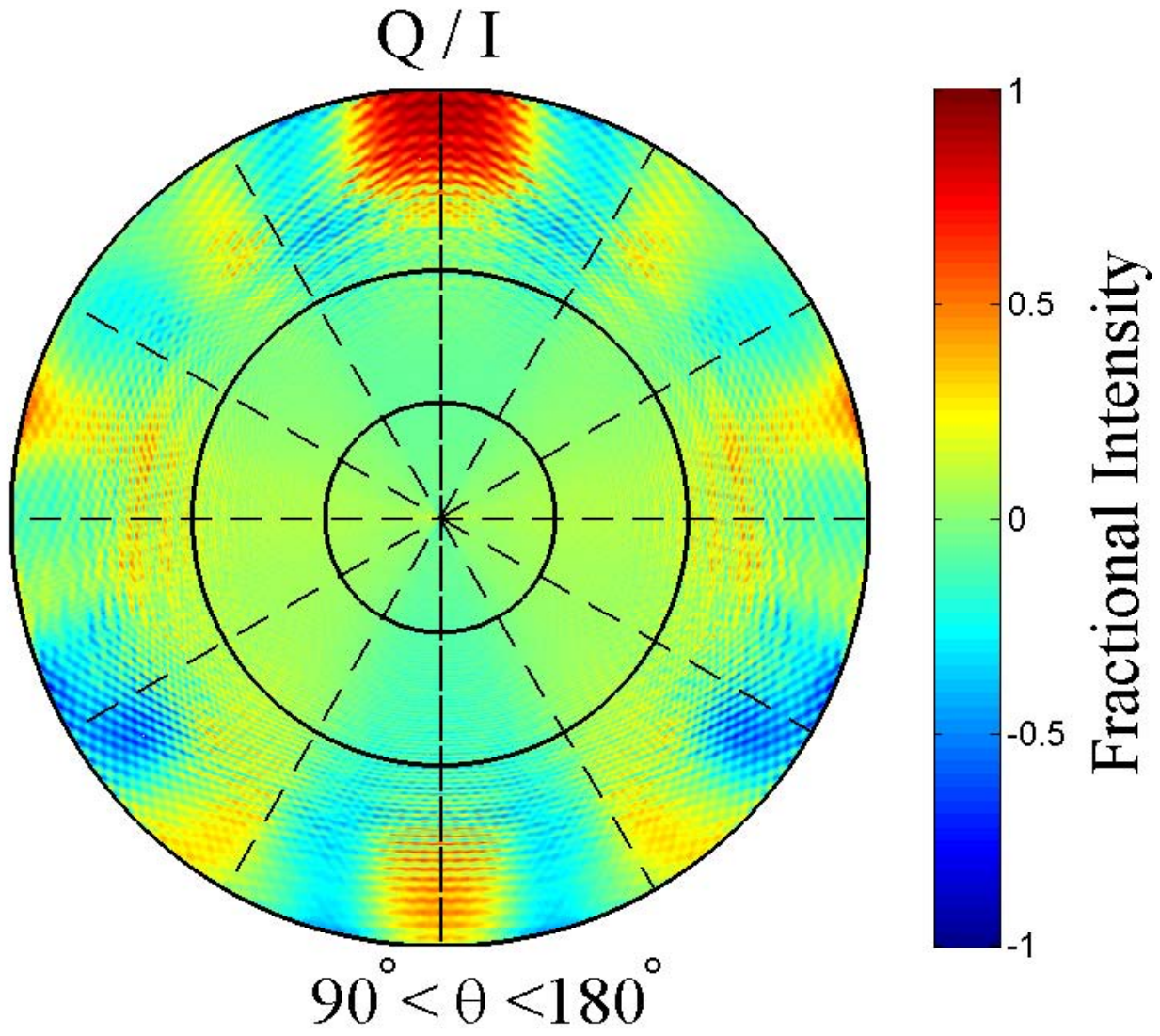}}
    \centerline{
    \includegraphics[bb = 195 150 598 470,width=0.27\textwidth,clip]
    {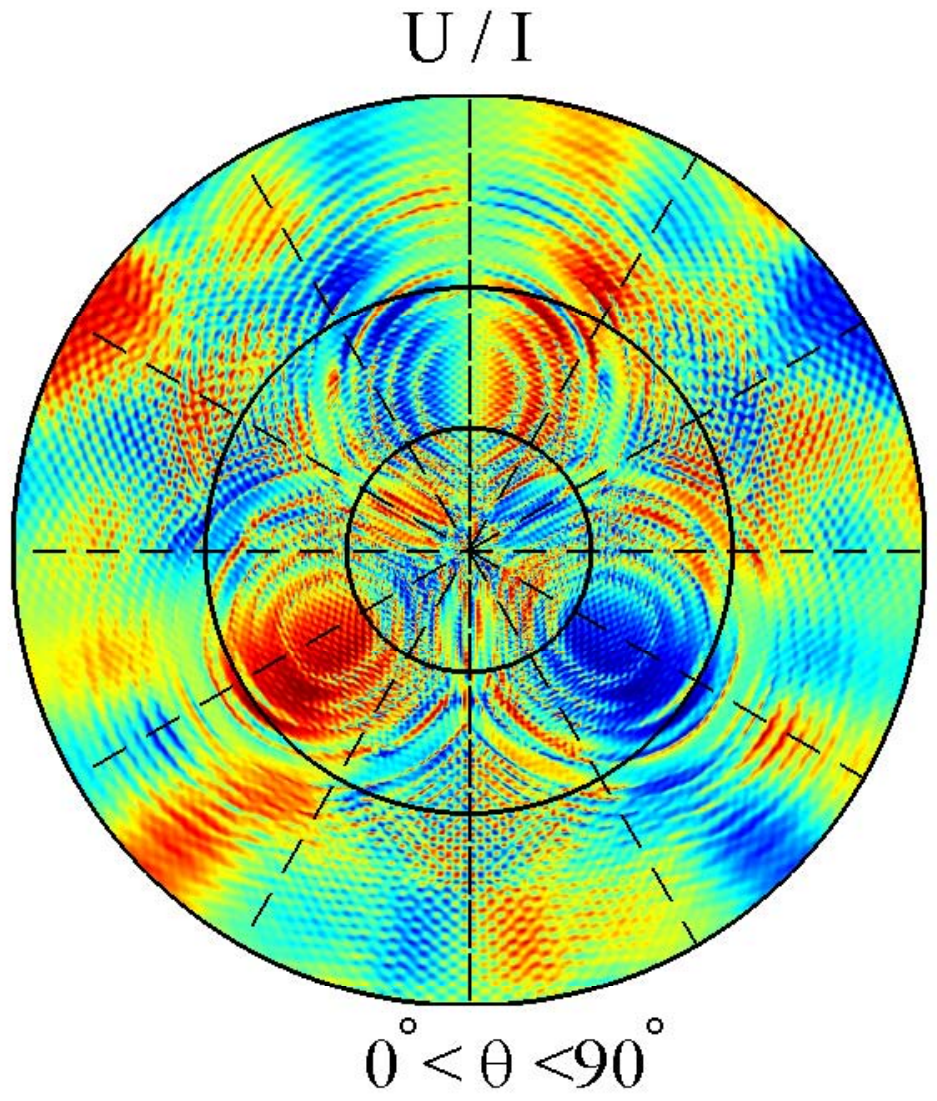}
    \includegraphics[bb = 195 150 598 470,width=0.27\textwidth,clip]
    {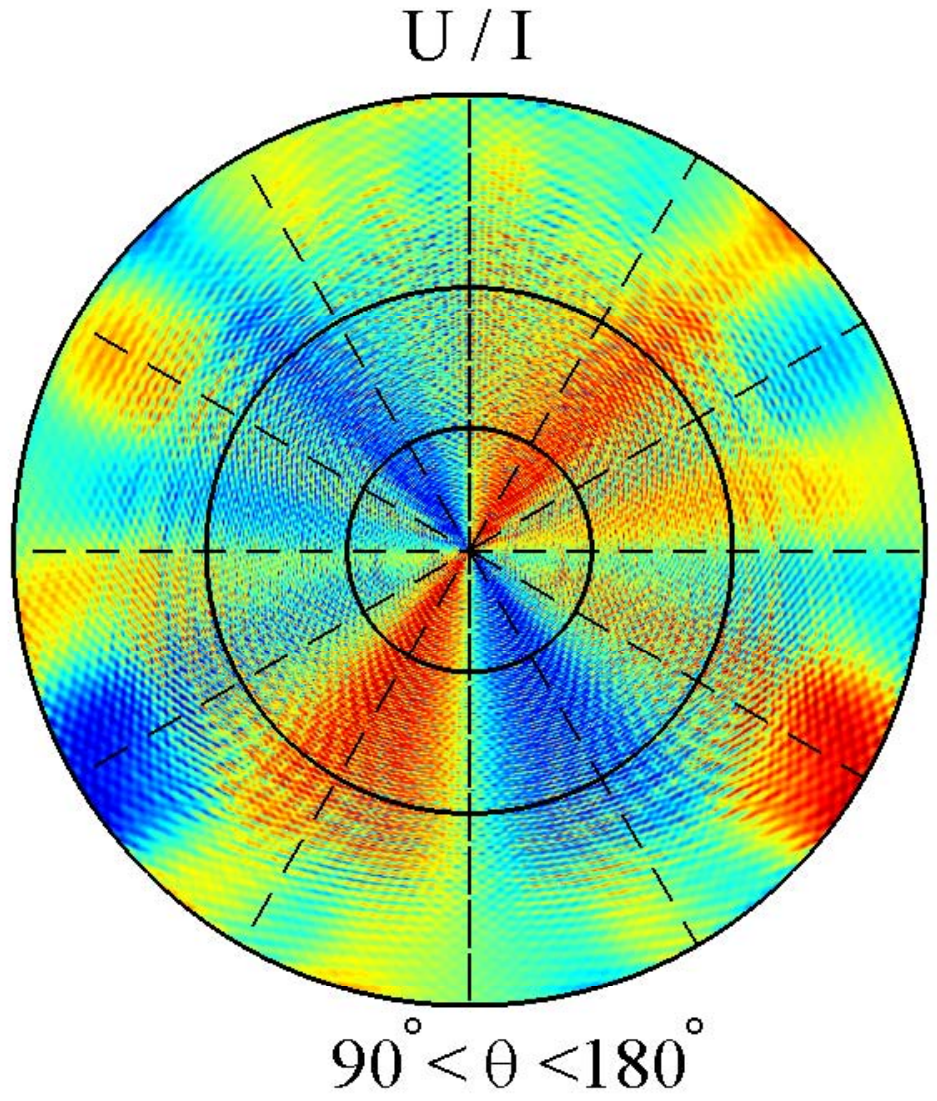}
    \includegraphics[bb = 195 150 598 470,width=0.27\textwidth,clip]
    {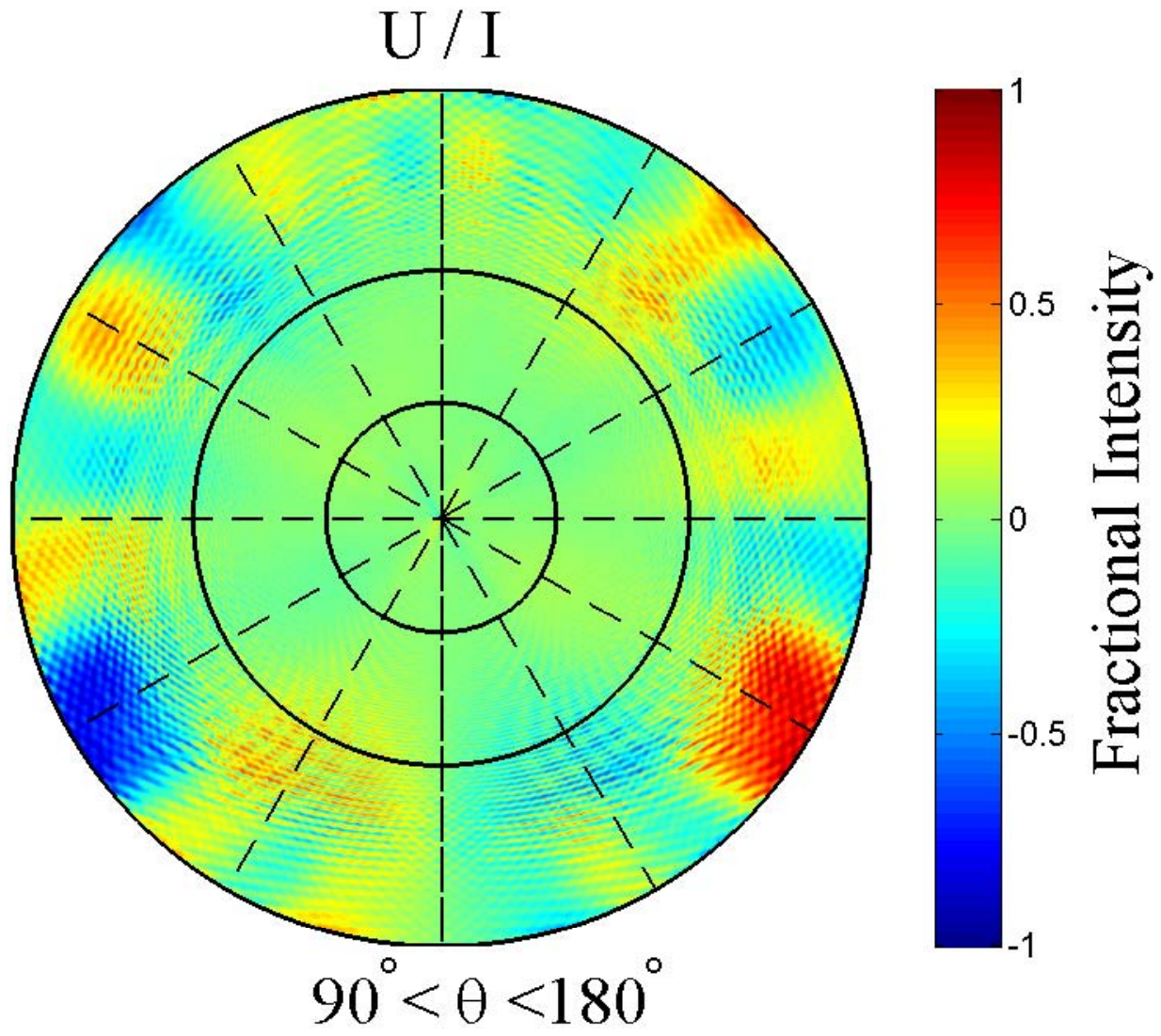}}
    \centerline{
    \includegraphics[bb = 195 150 598 470,width=0.27\textwidth,clip]
    {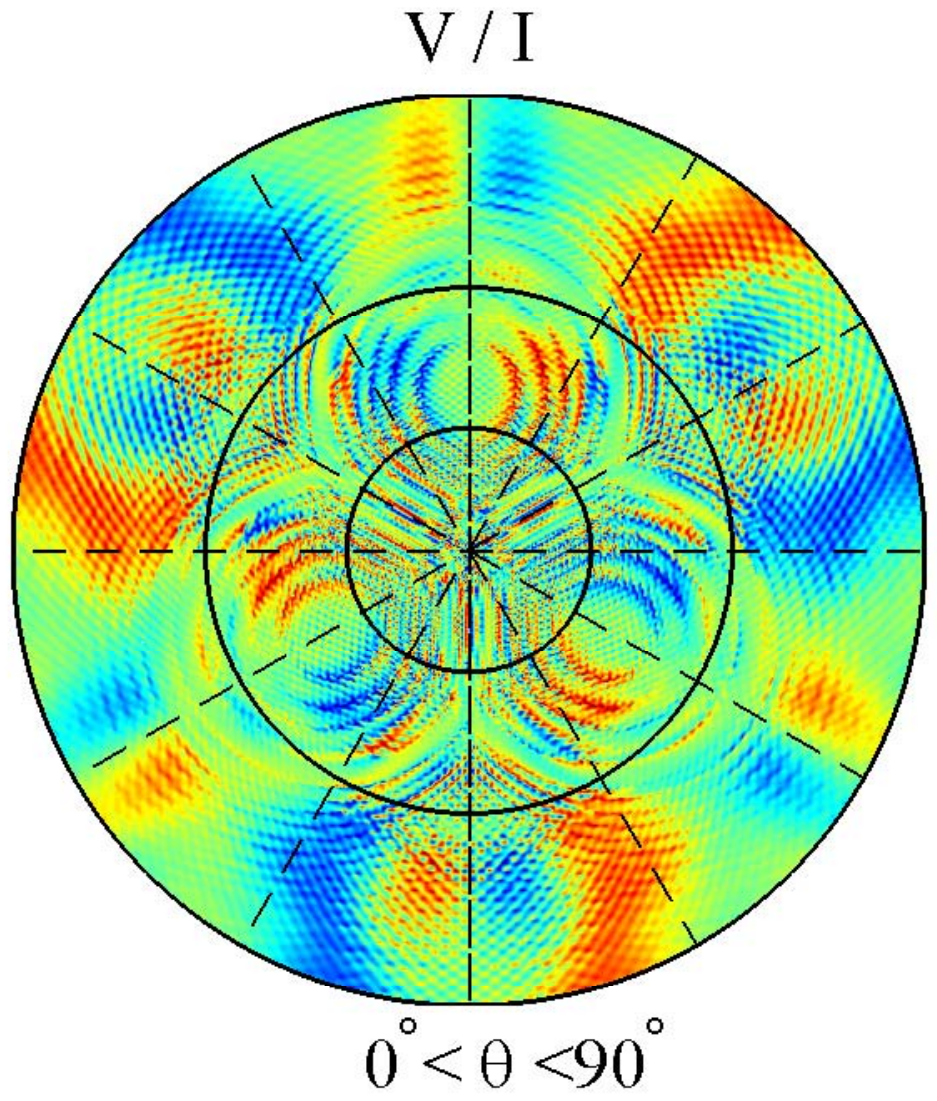}
    \includegraphics[bb = 195 150 598 470,width=0.27\textwidth,clip]
    {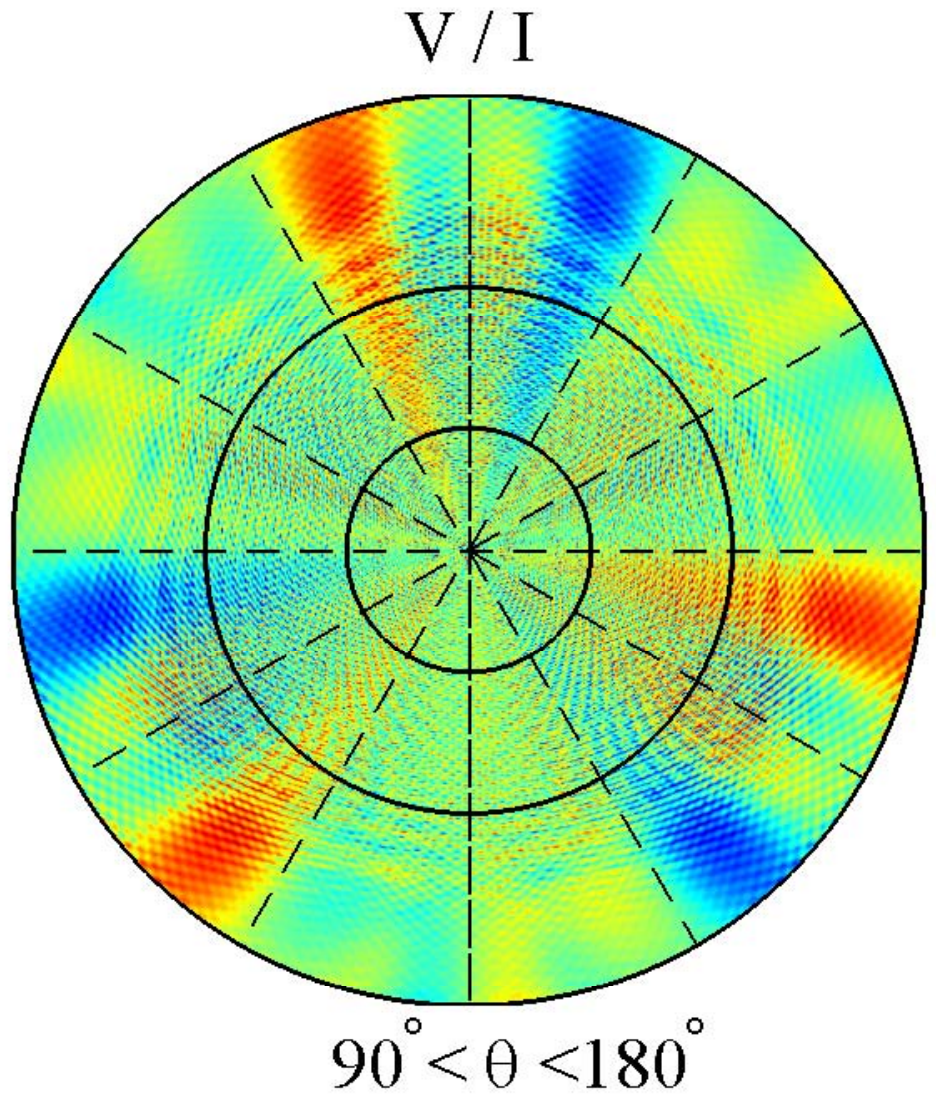}
    \includegraphics[bb = 195 150 598 470,width=0.27\textwidth,clip]
    {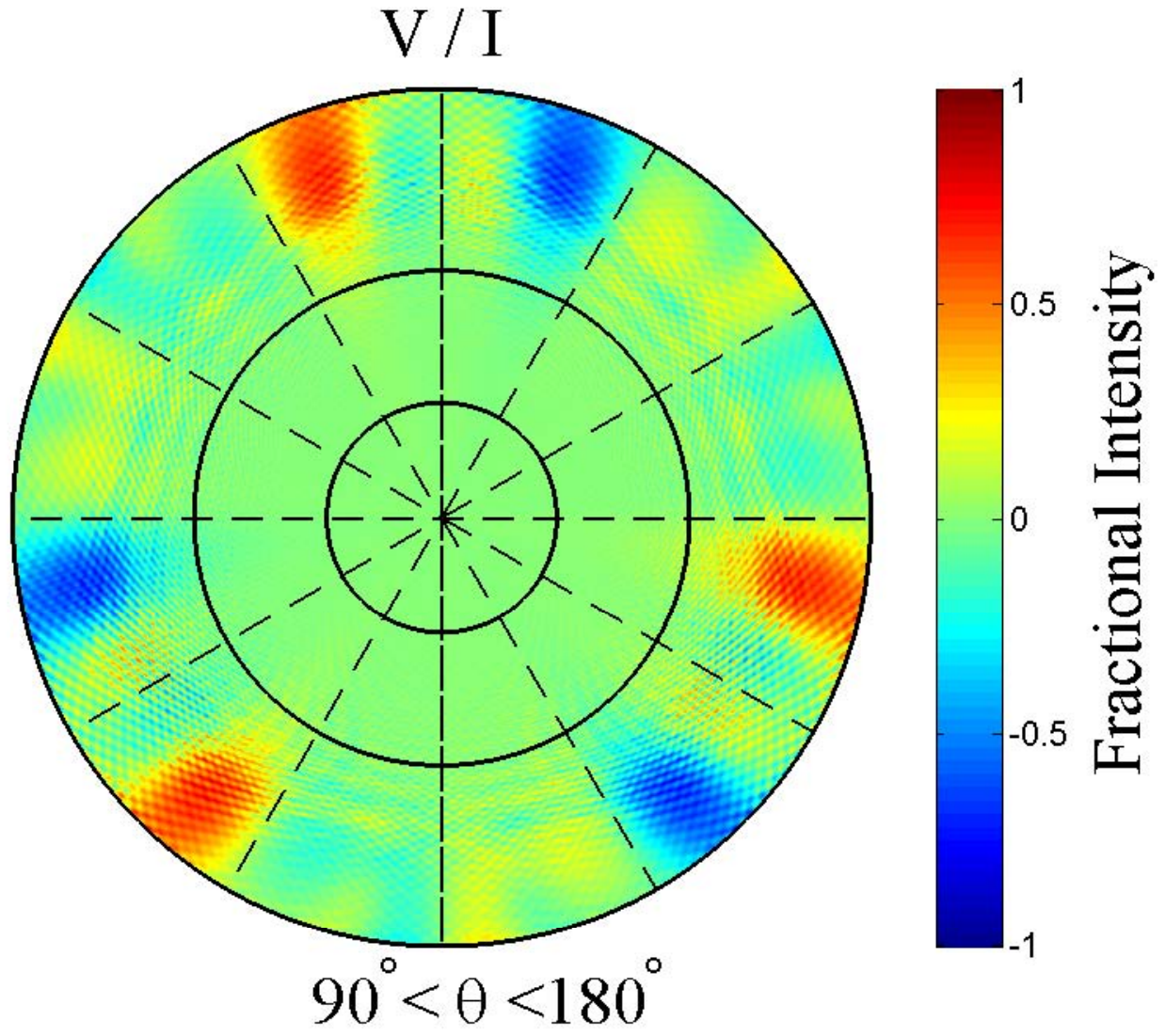}}
    \caption{Computed radiation patterns at 1400~MHz for Stokes $I$, $Q/I$,
       $U/I$, and $V/I$ shown for front and back hemispheres in
       stereographic projection in the coordinate system ${\theta},{\phi}$,
       where $\theta$ is the angle from boresight and $\phi$ is the
       circumferential angle. Units of $I$ are decibels relative to
       isotropic. Units of $Q/I$, $U/I$ and $V/I$ are fractional intensity. The left
       two columns show the front and back hemispheres of the radiation pattern
       from a reflector with surface errors (see text for details) but no
       leakage through the mesh. The right column shows the back hemisphere of
       the radiation patterns when leakage is added. The three lines superimposed
       on the forward $I$ pattern show the orientation of the feed-support struts;
       this figure is not an accurate picture of the telescope.}
  \label{rp}
\end{figure*}

\begin{figure}
   \centerline{\includegraphics[width=0.48\textwidth]
   {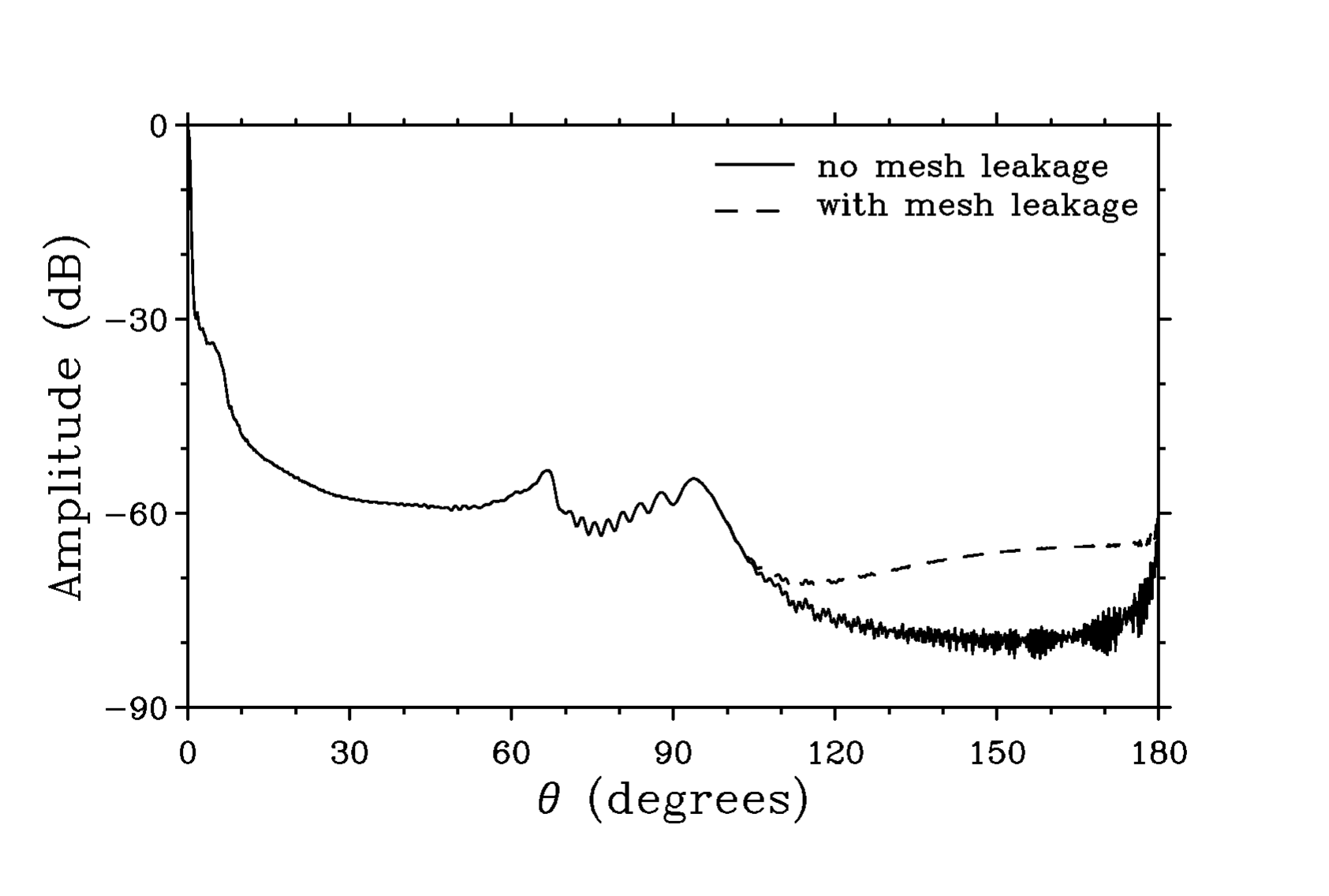}}
   \caption{Radial average of the computed radiation pattern at 1400~MHz as a function of the polar angle $\theta$, shown with and without radiation leaking through the mesh. The peak at $\theta\approx68^\circ$ arises from the scatter cones, and the peak at $\theta\approx95^\circ$ from spillover. Mesh leakage affects the pattern only for $\theta > 100^\circ$.}
   \label{radial}
\end{figure}

\begin{figure}
   \centerline{\includegraphics[width=0.3\textwidth,clip]{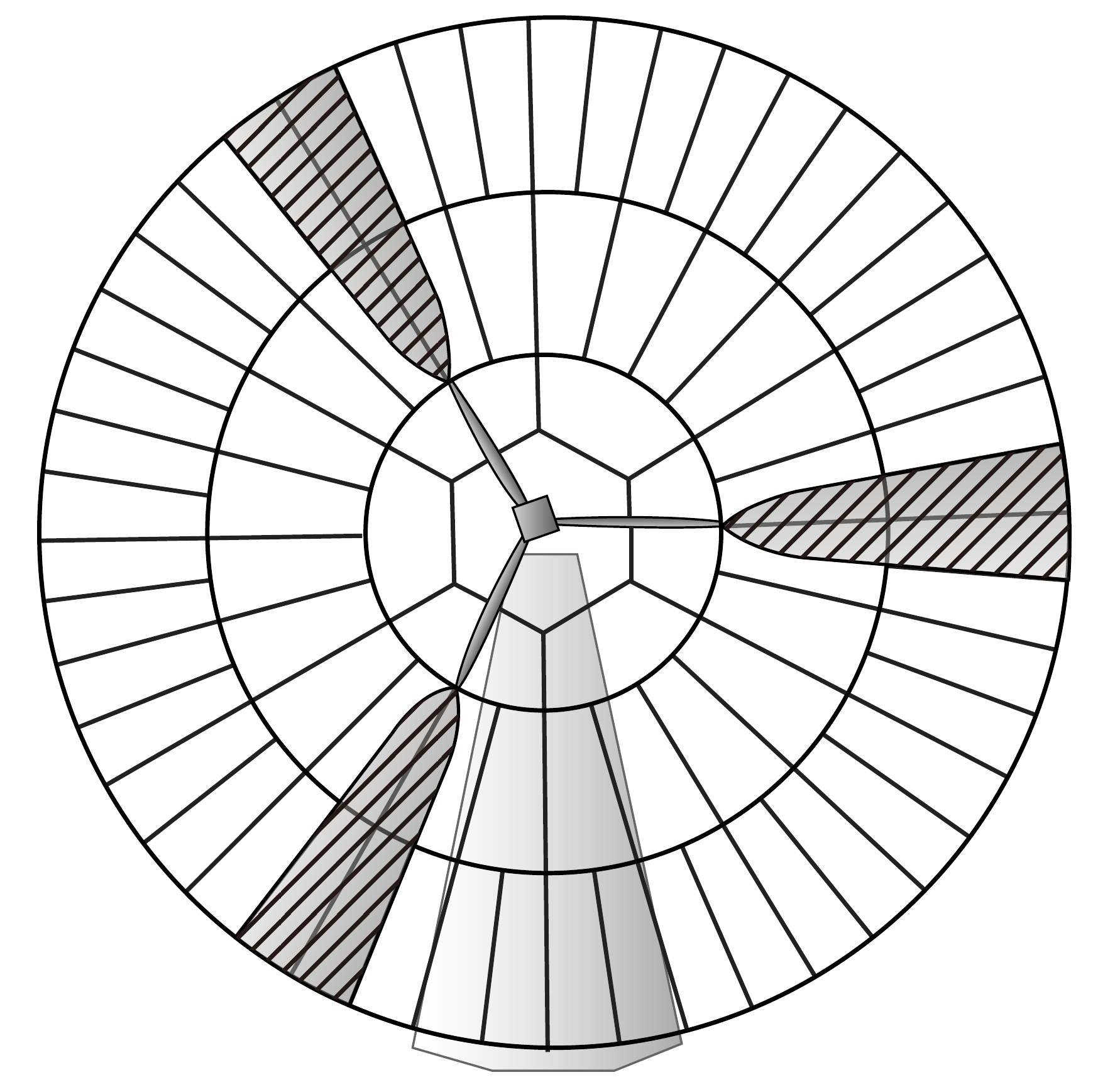}}
   \caption{Face-on view of the telescope. The parts of the reflector
   shadowed by the feed-support struts are hatched. The pale shaded
   structure behind the reflector is the support tower.}
   \label{shadows}
\end{figure}

Figure~\ref{rp} shows the results of GRASP-10 calculations using the technique described above. The pattern denoted $I$ is the familiar radiation pattern of antenna engineering. The three other patterns are generated by computing the radiated field in terms of $Q$, $U$ and $V$ and dividing each by $I$. The $Q/I$, $U/I$, and $V/I$ patterns indicate the conversion of unpolarized emission into apparently polarized signal. They indicate a potential error term that must be corrected.\footnote{While this is the dominant problem in
polarimetry of the linearly polarized Galactic emission, we note that a similar technique could be used to compute the conversion of any Stokes parameter into any other, for example $Q$ or $U$ into $V$ (the conversion of linearly polarized
emission into apparently circularly polarized signal). If a linearly polarized radiation pattern is calculated, the transfer of signal into $V$ can easily be determined. Such a calculation would be useful for observations of Zeeman
splitting of the H\thinspace\scriptsize{I} line.} They show the response of a polarization receiver attached to the telescope to a randomly polarized sky. Such a sky should produce a response in $I$ only, and the responses in $Q/I$, $U/I$ and
$V/I$ constitute the instrumental polarization whose effects we must correct for if we are to obtain a true representation of the sky.

The calculations shown in the left two columns of Figure~\ref{rp} include the effects of the three feed-support struts and include departure of the surface from a perfect paraboloid, but they do not include leakage through the mesh that forms the reflector surface.  The $I$ radiation pattern is dominated by the main beam, the ring of {\it{spillover}} emission, and the {\it{scatter cones}}. Spillover is direct feed radiation that is not intercepted by the reflector. It begins at the reflector rim (${\theta}={80^{\circ}}$) in the front hemisphere and continues to dominate in the back hemisphere, with some redistribution from diffraction at the reflector rim. The scatter cones \citep{land91} are the three circles that intersect on the main beam and dominate sidelobes in the forward hemisphere. They are generated by interaction of the plane wave rising from the paraboloidal surface of the reflector with the feed-support struts. The plane wave induces currents in each strut, and the strut becomes a travelling-wave antenna, radiating into a cone whose opening angle ($68^{\circ}$) is twice the angle between the strut and the telescope axis. The dominance of the spillover and the scatter cones can be appreciated from the plot in Figure~\ref{radial} which shows the radial average of the radiation pattern.

Shadowing by the struts is strong in this telescope (see Figure~\ref{shadows}). This translates into the $I$ pattern, producing a variation with $\phi$ up to $\pm$5 dB in the range ${60^{\circ}}<{\theta}<{100^{\circ}}$.

Both the spillover lobes and the scatter cones are strong features of the $Q/I$, $U/I$, and $V/I$ images. In other words these regions of the radiation pattern are strongly polarized. This is expected: the rim of the reflector and the feed-support struts are linear structures (certainly they are linear on the scale of the wavelength, ${\lambda}\approx{20\thinspace{\rm{cm}}}$); each will have markedly different response to radiation parallel to the linear structure and radiation orthogonal to it. It is also clear from Figure~\ref{rp} that the struts have a strong interaction with the spillover lobes. While the principal source of the spillover lobes is direct radiation from the feed, the spherical wave from the feed has a strong interaction with the struts.

The calculated $I$ response shows that the radiation scattered by the struts is quite concentrated into the scatter cones. Sidelobes which follow the form of the scatter cones are visible in other parts of the radiation pattern, but at a low level. However, in $Q/I$, $U/I$, and $V/I$ the effects of the scatter cones heavily modulate the radiation pattern over almost the entire front hemisphere. Figure~\ref{rp} tells us that understanding instrumental polarization is a matter of understanding how the spillover lobes and the scatter cones interact with the telescope environment.

All the patterns, $Q/I$, $U/I$, and $V/I$, reach values close to 100\%, telling us that there are regions of the radiation pattern that are almost completely polarized. The $Q/I$ and $U/I$ patterns reach these values at ${\theta}\approx{45^{\circ}}$ within the scatter cones and at certain azimuths in the spillover region. Peak values of $V/I$ are found at large $\theta$, but the dominant effects will come from the spillover regions.

We expect the $Q/I$ and $U/I$ patterns to resemble each other but with a $45^{\circ}$ rotation between them. This is evident in the rear hemisphere, but in the front hemisphere the influence of the $120^{\circ}$ azimuthal spacing of the struts has an equal effect.

Computation of the radiation pattern from a perfect reflector serves to illustrate the major influences at work, but the real antenna has a surface that deviates from the perfect paraboloidal shape. The surface accuracy was established by \citet{higg00} as ${\delta}={0.45}\thinspace{\rm{cm}}$\thinspace{rms} from measurements of the aperture efficiency at different wavelengths. We included surface roughness of this magnitude in the GRASP-10 calculations. Surface roughness decreases the gain, $G$, of the antenna, and we compared the gain reduction factor, $k_G$, derived from our GRASP-10 results with that expected from the formula of \citet{ruze66}, ${k_G}={e^{{-(4{\pi}{\delta}/{\lambda})}^2}}$. The results of GRASP-10 calculations agree with the predictions of the Ruze formula within 3\% at all frequencies. The gain reduction, ${\Delta}G$, from surface roughness at 1400~MHz is 6.4\%.

The mesh surface is also partly transparent at ${\lambda}\approx{20\thinspace{\rm{cm}}}$,  and allows some feed radiation to pass directly into the rear hemisphere (and, equivalently, some ground radiation to reach the feed through the mesh). We calculated the transparency of the mesh from its dimensions using the empirical formula given by \cite{mumf61}. The power leaking through the mesh ranges from 0.3\% of the incident power at 1250~MHz to 0.7\% at 1750~MHz.  We calculated the transmission coeffient, $\mu$, and used it to compute a radiation pattern incorporating the leakage radiation with the GRASP-10 output, combined vectorially at the appropriate levels. The leakage signal is feed radiation, attenuated, with its polarization state unaltered. The leakage through the mesh swamps the diffracted fields in the rear hemisphere, and the fractional polarization becomes quite low when leakage is included. Leakage through gaps in the reflector surface was not considered.

Radiation patterns at 1400~MHz showing all contributing factors are presented in Figure~\ref{rp}, and a radial average of the Stokes $I$ pattern is shown in
Figure~\ref{radial}. In all subsequent discussion we use these radiation patterns and the similar patterns at other frequencies.

\section{Results --- Aperture and Beam Efficiencies}

\subsection{Aperture Efficiency and Beam Efficiency from GRASP-10}
\label{gain}

Figure~\ref{eta-beta} shows the calculated aperture efficiency, $\eta_A$, and beam efficiency, $\eta_B$, of the telescope. Fitted curves are shown for both parameters, confined to the frequency range of the GMIMS observations, 1280 to 1750~MHz.

To calculate $\eta_B$ we used the following procedure. The half-power beamwidth, $\Theta(\nu)$, of the telescope at frequency $\nu$ varies from $\sim 40'$ at 1270~MHz to $\sim 30'$ at 1740~MHz. We used measured values of $\Theta(\nu)$ and assumed equal $E$- and $H$- plane widths at all $\nu$ (a very good approximation). At each frequency the solid angle of a Gaussian beam was calculated as ${{\Omega}_B({\nu})}={1.13\thinspace{\Theta(\nu)}^2}$. Total antenna solid angle, $\Omega(\nu)$, was derived from the GRASP-10 calculations, and main beam efficiency was calculated as ${\Omega_B}(\nu)/{\Omega}(\nu)$.

Much more complicated determinations of the solid angle of the beam have been used in the past. Some authors have integrated the main beam to the first null in the response. We tried this, identifying the first null in the GRASP-10 patterns, but there were fluctuations in the radius to the first null from one frequency to the next, and these fluctuations led to rapid and unrealistic variation in $\eta_B$ with frequency, so we rejected this option. Other authors have used the ``full-beam brightness temperature'', integrating a measured beam out to some radius (e.g. \citealp{reic82}). This seemed to us an arbitrary procedure, and we could think of no logical way to either choose the radius of the full beam, or alter that radius with frequency. With the wide bandwidth of the GMIMS dataset there are investigations (e.g. spectral index determination) which will compare the data at two frequencies within the dataset. We considered it important to keep consistent processing from one frequency to the next.

It might also be argued that the antenna beams are not Gaussian, and that a better approximation to the beamshape might be found from some other mathematical function. While that is probably true, we note that the GMIMS data will be subject to further processing which will assume that the observing beams {\it{are}} Gaussian. For example, in preparation for Rotation Measure Synthesis \citep{bren05} all data are convolved to a common beamwidth, that at the lowest frequency of the observations.

The small dip in value of $\eta_A$ and $\eta_B$ at 1550~MHz appears to arise from a property of the feed. At that frequency the feed radiation pattern has a distinctly flat top (see Figure~\ref{primary}). This dip in aperture efficiency cannot be seen in our measurements (Figure~\ref{eta-meas}), and we have not investigated it any further. The fitted curves in Figure~\ref{eta-beta} are fits to all data points, including the value at 1550~MHz, but only a simple quadratic function has been fitted and this cannot include the dip.

Values of $\eta_A$ and $\eta_B$ were needed for reduction of the GMIMS data. The survey was calibrated (in janskys) with daily observations of strong sources whose flux densities were well known. Values of $\eta_A$ were used to convert observational data from janskys to antenna temperature. We used the curve fitted to the GRASP-10 values of $\eta_A$ shown in Figure~\ref{eta-beta}, adjusted to fit our measurements of $\eta_A$ (see Section~\ref{meas} and Figure~\ref{eta-meas}). Finally, the data were converted to main-beam brightness temperature using the values of $\eta_B$.

\begin{figure}
   \centerline{\includegraphics[width=0.48\textwidth]{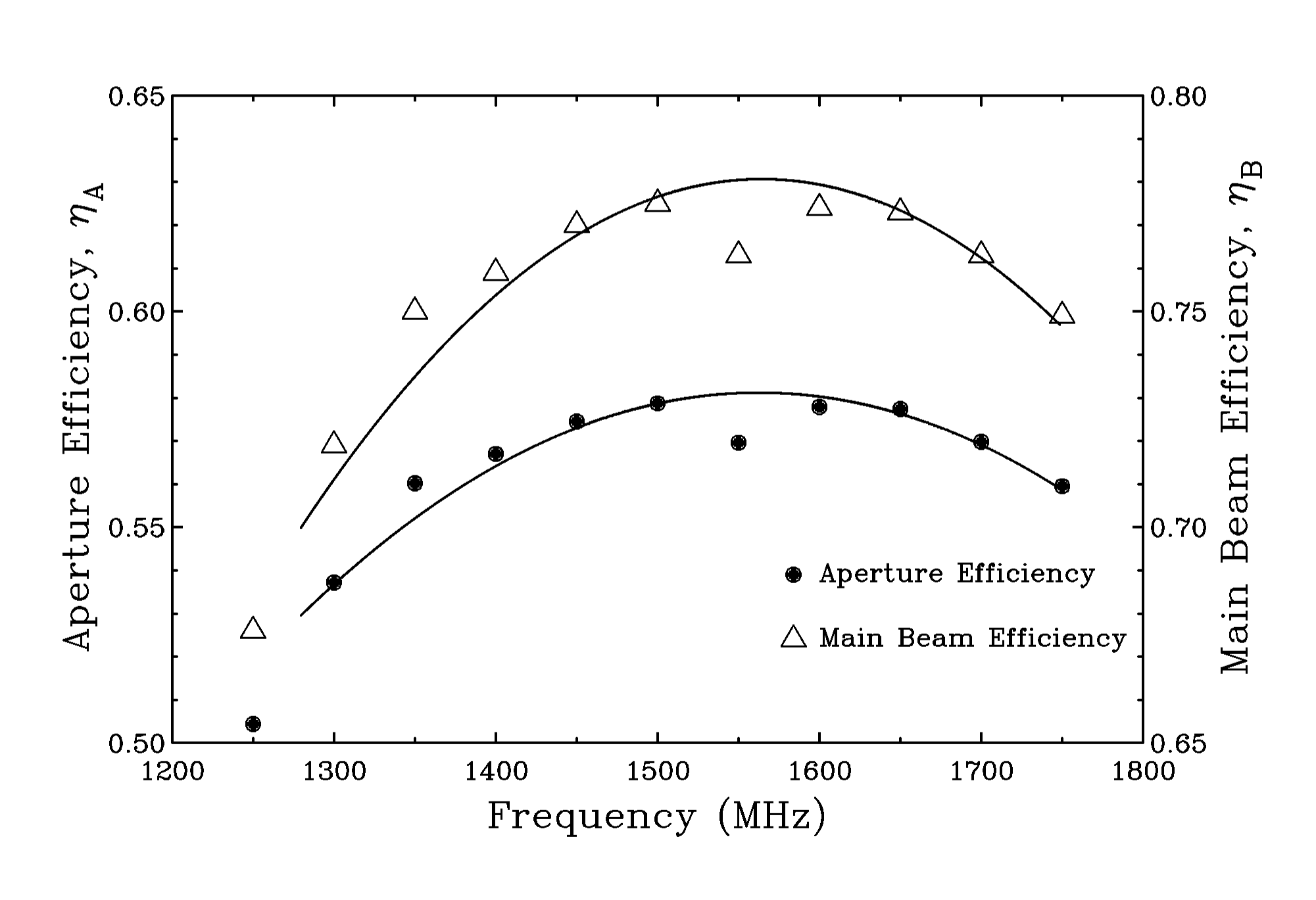}}
      \caption{Aperture efficiency, $\eta_A$, calculated with GRASP-10 and main beam
   efficiency, $\eta_B$, calculated as described in the text. Surface roughness
   and leakage have been included in the calculations. Curves have been fitted
   to the data over the frequency range of the GMIMS data (the points at 1250
   MHz were excluded from the fits).}
     \label{eta-beta}
\end{figure}

\subsection{Aperture Efficiency from Geometrical Optics and Ray Tracing}

We used a second approach to the determination of aperture efficiency, using much simpler techniques in order to provide a check of the GRASP-10 calculations.  For this discussion it is convenient to subdivide the aperture efficiency into a number of efficiencies, each of which can be separately evaluated. Thus
\begin{equation}
{\eta_A}={{\eta_i}\,{\cdot}\,{\eta_s}\,{\cdot}\,{\eta_e}
\,{\cdot}\,{\eta_l}\,{\cdot}\,{\eta_b}} \thinspace.
\label{subdiv}
\end{equation}
We now discuss the individual terms on the right-hand side of
equation~\ref{subdiv} (using  the language of transmitting and receiving
interchangeably).

\begin{itemize}

\item{Two efficiencies--- $\eta_i$, the illumination efficiency, and $\eta_s$, the spillover efficiency--- reflect the compromise inherent in the design of the feed between the efficient use of the reflector and loss of power into the spillover region. $\eta_i$ is the integrated power across the aperture relative to the power in a uniformly illuminated aperture. $\eta_s$ is the fraction of power that is intercepted by the reflector; the fraction $1-\eta_s$ goes beyond the edge of the reflector and forms the spillover lobes. High levels of $\eta_i$ usually mean low levels of $\eta_s$ and vice versa.}

\item{$\eta_e$ is the surface efficiency. Small surface deflections produce phase errors that reduce antenna gain by $1-\eta_e$.}

\item{$\eta_l$ is the leakage efficiency. This reflects loss of a fraction of $1-\eta_l$ through the mesh surface of the reflector or through gaps between reflector panels.}

\item{$\eta_b$ is the blockage efficiency. The focus equipment and the struts that support it obviously block the aperture, so ${\eta_b}<{1}$. Two effects must be considered: the struts block the incoming plane wave, but they also shadow part of the reflector surface so that the feed cannot receive signal from there. These two effects are known as plane-wave blockage and spherical-wave blockage respectively and we separately evaluate efficiencies ${\eta}_{\rm pw}$ and ${\eta}_{\rm sph}$.}

\end{itemize}

We calculated $\eta_i$, $\eta_s$, $\eta_{\rm pw}$, and $\eta_{\rm sph}$  based on geometrical optics and ray tracing. The calculations were made using a program written by L.A. Higgs \citep{hk00}. The inputs to the program were the physical dimensions of the antenna and the radiation patterns of the feed. We used the patterns calculated by CST (see Section~\ref{feed}), except that the $E$- and $H$-plane patterns were averaged to provide a circularly symmetric illumination of the reflector (the actual patterns come close to this).

Assuming an unblocked aperture, the program calculates the field on the aperture, taking into account the shape of the reflector. The level of feed radiation drops off towards the edge of the reflector (see Figure~\ref{primary}) but there is an additional ``free-space attenuation'' \citep{baar07} which further reduces illumination at the edge of the aperture. The radiation pattern of the feed is measured (or calculated) on a spherical surface centered on the feed, but the outer edges of the aperture plane are further from the focus than the center. This adds an extra taper
\begin{equation}
{\zeta} = {1 + \Biggl({\frac{rD}{4f}}\Biggr)^2} \thinspace,
\end{equation}
where $f$ is the focal length, $D$ is the antenna diameter, and $0<r<1$ is radial distance in the aperture \citep{baar07}. The free-space attenuation is stronger for deeper reflectors; for the Galt Telescope, where ${f/D}=0.298$, it amounts to an additional taper of 4.6~dB at the edge of the aperture beyond that provided by the feed radiation pattern.

The illumination efficiency, $\eta_i$, was calculated from the numerical data for the field distribution on the aperture using the equation
\begin{equation}
{\eta_i} =
{\frac{\biggl(\int{F{\thinspace}dA}\biggr)^2}{{A_p}\int{{F^2}{\thinspace}dA}}} \thinspace,
\label{etai}
\end{equation}
where $F$ is the power distribution in the aperture, including the free-space attenuation, and the integrations extend over the aperture whose physical area is $A_p$ \citep{coll85}.

Calculation of $\eta_s$ is simply an integration of the power radiated by the feed that misses the reflector.

Plane-wave blockage is derived by calculating the geometrical shadow of the central focus equipment and the feed-support struts on the aperture under plane-wave illumination. The tapered illumination of the aperture is taken into account.

The program derives spherical wave blockage by calculating the geometric shadows of the struts on the aperture, taking into account their cigar shape. We use the
relationship from \citet{lamb86} to calculate the blockage
\begin{equation}
{B}= {\frac{ \int_{\rm blockage}{E{\thinspace}dA}}{\int_{\rm aperture}{E{\thinspace}dA}}} \thinspace,
\label{sph-block}
\end{equation}
where $E$ is the amplitude of the aperture field and $dA$ is an area element.
Then
\begin{equation}
{{\eta}_{\rm sph}}={(1-B)^2} \thinspace.
\end{equation}

Effects of surface roughness and of leakage through the mesh, ${\eta}_e$ and ${\eta}_l$, were evaluated as described in Section~\ref{radpa}.

The results of these calculations are summarized in Table~1. Results are tabulated to four significant figures; four-digit precision is not justified by the accuracy of the input data but is preserved so that the gradual trends with frequency remain clear to the reader.

\begin{table*}
\begin{center}
    \small
    \begin{tabular}{|c|c|c|c|c|c|c||c|c|}
          \hline
            frequ- & illumin- & spill- & surf- & leak- & plane & spherical & ray & GRASP \\
            ency & ation & over & ace & age & wave & wave & tracing & efficiency \\
            (MHz) & ${\eta}_I$ & ${\eta}_s$ & ${\eta}_e$ & ${\eta}_l$ & ${\eta}_{\rm pw}$ &
            ${\eta}_{\rm sph}$ & efficiency &  \\
                      \hline
            	  \hline
            1250 & 0.6807 & 0.9323 & 0.9460 & 0.9967 & 0.9554 & 0.8032 & 0.4592 & 0.5044 \\
                      \hline
            1300 & 0.6978 & 0.9533 & 0.9417 & 0.9964 & 0.9598 & 0.8701 & 0.5213 & 0.5372 \\
                      \hline
            1350 & 0.7191 & 0.9602 & 0.9373 & 0.9961 & 0.9628 & 0.8673 & 0.5383 & 0.5602 \\
                      \hline
            1400 & 0.7290 & 0.9602 & 0.9327 & 0.9958 & 0.9641 & 0.8660 & 0.5428 & 0.5670 \\
                      \hline
            1450 & 0.7338 & 0.9591 & 0.9280 & 0.9955 & 0.9646 & 0.8655 & 0.5428 & 0.5745 \\
                      \hline
            1500 & 0.7359 & 0.9573 & 0.9237 & 0.9952 & 0.9650 & 0.8651 & 0.5407 & 0.5787 \\
                      \hline
            1550 & 0.7215 & 0.9551 & 0.9182 & 0.9949 & 0.9664 & 0.8668 & 0.5273 & 0.5696 \\
                      \hline
            1600 & 0.7295 & 0.9681 & 0.9131 & 0.9946 & 0.9669 & 0.8656 & 0.5368 & 0.5779 \\
                      \hline
            1650 & 0.7252 & 0.9677 & 0.9078 & 0.9942 & 0.9640 & 0.8622 & 0.5289 & 0.5774 \\
                      \hline
            1700 & 0.7128 & 0.9694 & 0.9024 & 0.9939 & 0.9619 & 0.8679 & 0.5174 & 0.5698 \\
                      \hline
            1750 & 0.6970 & 0.9717 & 0.8969 & 0.9935 & 0.9592 & 0.8701 & 0.5037 & 0.5595 \\
          \hline
          \hline
        \end{tabular}
      \end{center}
\begin{center}
{\bf{Table 1:}} Contributions to aperture efficiency, and a comparison of ray tracing and GRASP-10 calculations.
\end{center}
\end{table*}

Values of $\eta_A$ obtained by the two methods are compared in Figure~\ref{eta-comp}. The ray-tracing results are below the GRASP-10 results by 1\% to 7\%, a fair agreement considering  the simplicity of the ray-tracing approach. We conclude that the ray-tracing technique can be used to determine aperture efficiency of a reflector antenna with useful accuracy. Both GRASP-10 and ray-tracing reveal a dip in $\eta_A$ at $\sim$1550~MHz. Since both techniques yield a similar result, we can trace the cause to the feed radiation pattern that is calculated by CST. We see from Table 1 that the illumination efficiency, which depends entirely on the feed characteristics, is lower at 1550 MHz than at adjacent frequencies. However, as remarked above, our measurements do not show this dip. We have not investigated it further and used the fitted curves shown in Figure 5 to process the GMIMS data.

\begin{figure}
   \centerline{\includegraphics[width=0.48\textwidth]{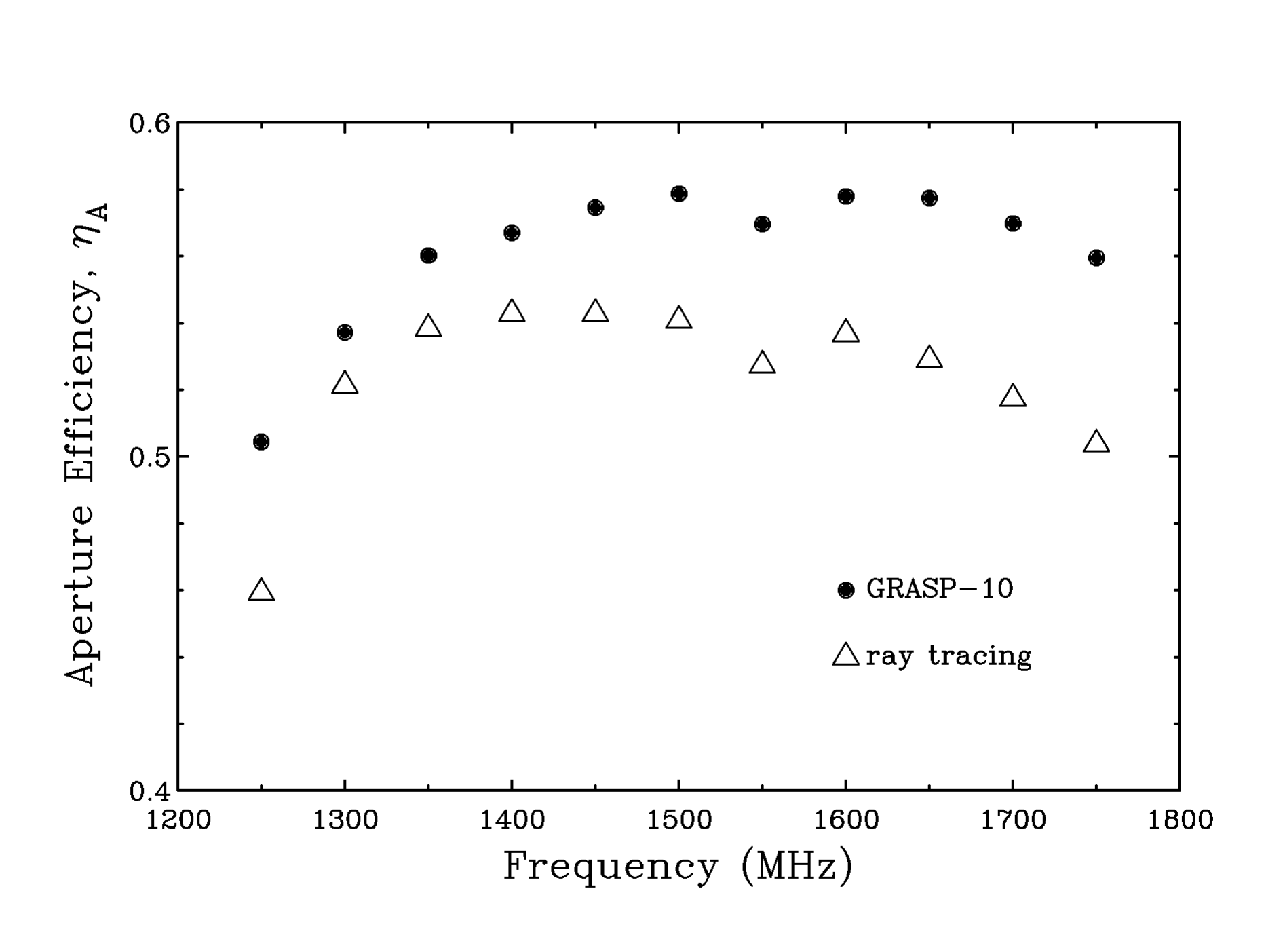}}
   \caption{Comparison of aperture efficiency, $\eta_A$, calculated with
   GRASP-10 and with ray tracing as a function of frequency.}
     \label{eta-comp}
\end{figure}

Table 1 shows that the dominant effects in determining the aperture efficiency are the illumination efficiency and the spherical wave blockage. Together these two factors restrict aperture efficiency to be below about 0.65; other effects are relatively minor. For the Galt Telescope spherical wave blockage is especially important because the feed-support struts are at an angle of only $34^{\circ}$ from the telescope axis while the reflector edge is ${\sim}80^{\circ}$ from the axis; the struts shadow a substantial part of the reflector from the spherical wave emanating from the feed, as is evident from Figure~\ref{shadows}.

\begin{figure}
   \centerline{\includegraphics[width=0.44\textwidth]{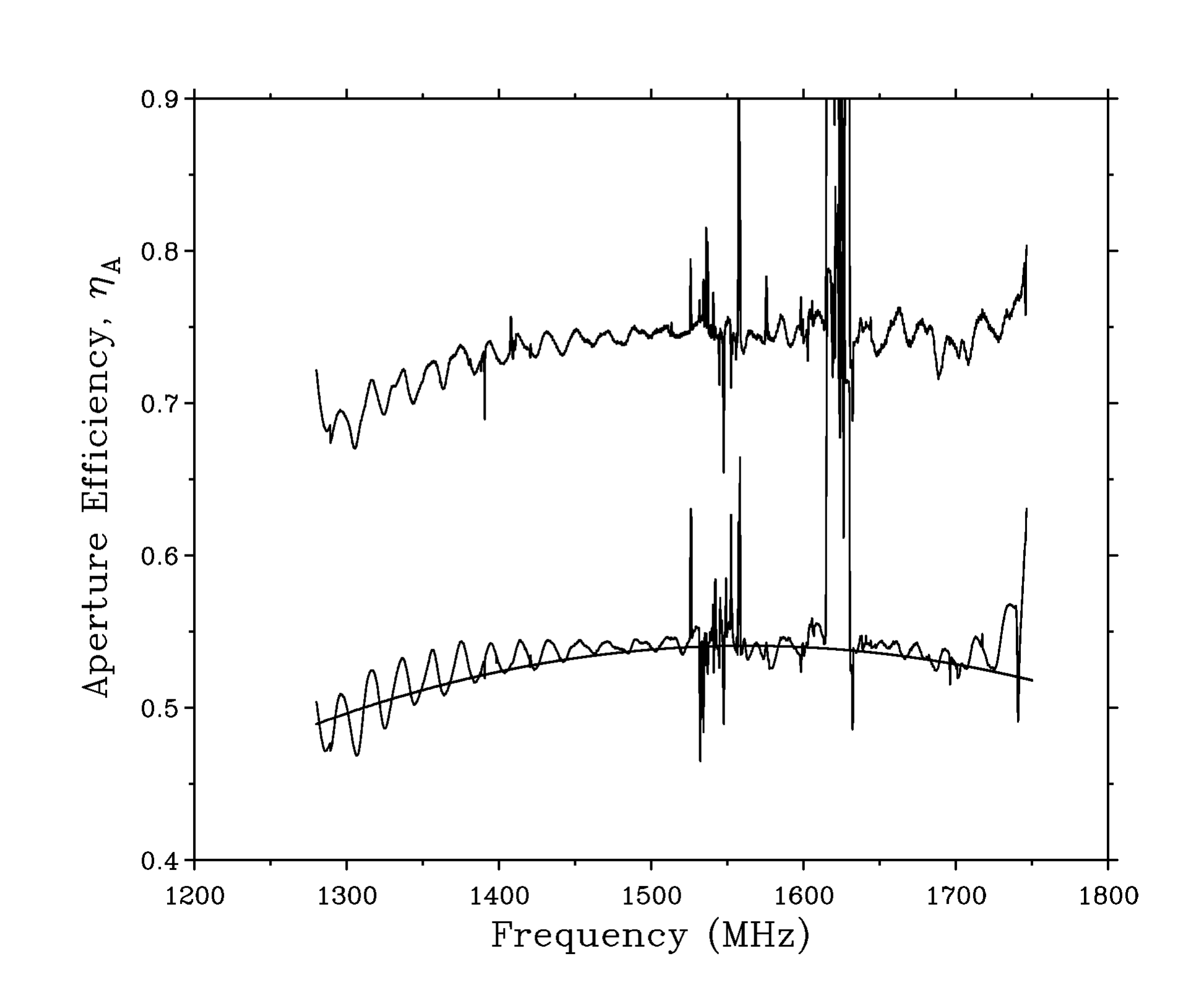}}
   \caption{Measured aperture efficiency, from observations of Cyg~A, in RHCP
   (lower curve) and LHCP (upper curve, displaced upwards by 0.2 for clarity).
   Effects of interfering signals are evident. The smooth curve is fitted to the
   GRASP-10 calculations, adjusted as described in the text.}

     \label{eta-meas}
\end{figure}

\section{Measurement of Aperture Efficiency}
\label{meas}

The GMIMS observations were calibrated with four strong sources, Cyg A, Tau A, Vir A, and Cas A. A self-consistent set of flux densities and spectral indices was generated by observations of these sources at the start of the survey \citep{woll10a}. Aperture efficiency of the telescope was subsequently measured from observations of Cyg A, assuming a flux density at 1.4~GHz of 1589~Jy and a spectral index of $-$1.07.

The receiver made all measurements of antenna temperature relative to a switched noise signal injected at 25 Hz into both LHCP and RHCP receiver channels. In a first measurement the amplitude of the noise signal in kelvins was established relative to coaxial resistive terminations at known temperatures, one at liquid nitrogen temperature, one at ambient temperature, and one in a temperature-controlled oven at ${\sim}100 ^{\circ}$C. This experiment did not include the feed and its associated waveguide components, so a second measurement was made which did include those components in the signal path. The high reference temperature was provided by a box filled with absorbing foam placed in front of and around the feed (all at ambient temperature). It was not possible to immerse this large box in liquid nitrogen, so the cold temperature was provided by driving the telescope to the zenith and using the cold sky together with informed estimates of the various contributions to the zenith antenna temperature.

The two measurements gave closely consistent results for the equivalent temperature of the injected noise signal over most of the band, with the exception of two frequency ranges where there were moderate mismatch problems in the feed. The second measurement, by including the feed in the signal path, overcame these difficulties, but had the slight disadvantage that the results were quite sensitive to the adopted value for the antenna temperature at the zenith. Our estimates of antenna temperature at the zenith allowed for ground emission received through the spillover sidelobes (see Section~\ref{spill}) and through the partly transparent mesh reflecting surface, atmospheric emission \citep{gibb86} and the brightness temperature of the sky (including Galactic and extragalactic contributions and the cosmic microwave background).

Figure~\ref{eta-meas} shows the aperture efficiency derived by observing Cyg~A. Results for the two hands of circular polarization, essentially two independent measurements, are shown individually.  The smooth curve in Figure~\ref{eta-meas} is based on the fit to the GRASP-10 values of $\eta_A$ shown in Figure~\ref{eta-beta}. The curve has been adjusted downwards by 0.29~dB (6.5\%) preserving its shape, to fit well with the Cyg~A determination. The observations provide strong confirmation that GRASP-10 has correctly computed the {\it{shape}} of the variation of $\eta_A$ with frequency, even if the absolute value requires some adjustment. The values of $\eta_A$ using ray tracing are a good fit to the measurements without any adjustment. We discuss the discrepancy between experimental and GRASP-10 values in Section~\ref{disc}.

The good agreement between the measured value of $\eta_A$ and calculations using GRASP-10 and using ray-tracing gives us strong confidence in the results. We adopt the adjusted GRASP-10 result, the smooth curve in Figure~\ref{eta-meas}, in processing the data from the GMIMS survey. We estimate the error in $\eta_A$ to be ${\pm}3\%$. The relative error in values of $\eta_A$ between any two frequencies in the band is less than this, about 1\%.

A strong ripple in $\eta_A$ is very obvious in Figure~\ref{eta-meas}. This is almost certainly an interaction of the feed with the reflector: the incident signal is not completely absorbed by the feed, and some is scattered from the feed and, after reflection from the vertex of the reflector, a delayed signal interferes with the prompt signal to form the ripple. The frequency separation of maxima is exactly what can be predicted from the focal length of the telescope.

The measured beamwidth of the telescope also shows a ripple with frequency (see Figure~1 of \citealp{woll10a}). The aperture efficiency is high at frequencies where the beamwidth is low and vice-versa, as expected. A change of $a\%$ in beamwidth will result in a change of $2a\%$ in gain and aperture efficiency, and this is (roughly) borne out by the measurements. Similar effects are reported by \citet{popp08}.

\citet{silv49} gives an expression for the reflection coefficient of the reflector-feed combination
\begin{equation}
{|\Gamma_r|}={\frac{{G_f}\thinspace{\lambda}}{4{\pi}f}} \thinspace,
\label{gamma_refl}
\end{equation}
where $G_f$ is the gain of the feed and $f$ is the focal length of the reflector. \citet{morr78} shows that the peak-to-peak magnitude of the ripple in the power seen by the receiver will be
\begin{equation}
{\frac{{\Delta}P}{P}}={4{\gamma}{\Gamma_r}} \thinspace,
\label{ripple}
\end{equation}
where $\gamma$, the scattering factor, describes scattering by the feed. \citet{morr78} evaluates $\gamma$ for a ${\rm{TE_{11}}}$ waveguide feed with a flange at its aperture. This is close to the feed on the Galt Telescope, but not exactly the situation. We have some difficulty deciding on the value of $\gamma$ because there are other things near the focus of the telescope that will aggravate the scattering, for example the feed-support structure and the receiver box. For a point source in the main beam (relevant to the data in Figure~\ref{eta-meas}) we take ${\gamma}\approx{1}$.

Equations \ref{gamma_refl} and \ref{ripple} predict that the ripple will be strongest at the low frequency end of the band, and the measurements in Figure~\ref{eta-meas} show this. Equation \ref{gamma_refl} is dominated by the wavelength, longest at the low end of the band, and feed gain is higher there than at band center. Both decrease to higher frequencies. However, in detail the match of measurement and prediction is not exact. At 1300~MHz equation~\ref{ripple} predicts ${\frac{{\Delta}P}{P}}={0.07}$; the measured value is 0.11. At 1500~MHz the predicted value is 0.05 and the measured value 0.02.

\citet{morr78} also points out that bandpass ripple will arise from extended emission in the main beam and from spillover radiation. The factor $\gamma$ in those two cases is different, and differs from the point-source value. We cannot confirm this from measurements, but note that amplitude ripples with the same characteristic frequency structure are certainly present in the GMIMS data, particularly at the low-frequency end of the passband.

\section{Spillover and Ground Radiation}
\label{spill}

Emission from the ground is a major factor in very sensitive radio astronomy systems because it contributes to system noise, and is often the largest contributor after the receiver itself.  The principal route by which ground emission reaches the receiver is by radiating directly into the feed, over the edge of the reflector, through the spillover sidelobes. The levels and forms of the spillover sidelobes are largely determined by the design of the feed: spillover radiation is feed radiation. The edge of the reflector is at an angle ${\sim}80^{\circ}$ from the feed boresight, and the spillover region extends from here past $90^{\circ}$ to an angle where the feed no longer has significant response. One can hardly expect that radiation at these angles, far from the axis of a circular waveguide feed, will have good polarization characteristics. Spillover levels will also be influenced by diffraction at the edge of the reflector. Again, diffraction from an edge is a very polarization-sensitive process \citep{bach86}. Ground radiation is scattered from the feed-support struts into the aperture and enters the receiver  \citep{ande91,land91}. The struts are long, relatively thin structures, and one would expect their scattering properties to be polarization dependent. For all these reasons, the paths through which ground emission enters the reflector are all likely to have poor polarization properties, which means that they are highly likely to convert the ground emission into spurious polarized signal.

Any discussion of the interaction of the radiation pattern with the ground is complicated by the fact that the ground radiation is itself partly polarized. We first discuss the properties of the ground as a polarized emitter and then consider the error introduced by ignoring the polarization and assuming that the ground signal is unpolarized.

\subsection{The Polarization of Ground Emission}
%\subsection{Interaction with Partially Polarized Ground Emission}
\label{ground-pol}

At frequencies around 1.5~GHz the ground is not a perfect absorber, it is a lossy dielectric, and its reflection and emission properties are polarization dependent. This is a well known effect at meter wavelengths where horizontal polarization is usually used for communications applications because such signals reflect strongly from the ground at grazing incidence and propagate reliably over large distances. In this discussion `horizontal' and `vertical' have particular meanings. The horizontally polarized component is normal to the propagation direction and parallel to the ground surface; the vertically polarized component is normal to the propagation direction and in the plane that is perpendicular to the ground. Note that these definitions are unique to each sidelobe of the telescope because they involve the look direction of that sidelobe.

We adopt a value of ground permittivity, ${\epsilon_r}={7}$, appropriate for 1.5~GHz \citep{hall79,itu00}.  We know that the ground is an emitter at our frequencies, so soil conductivity cannot be zero, but it is small under the dry soil conditions around the Observatory, typically ${\kappa}={0.001}\thinspace{\rm{S}}\thinspace{\rm{m}}^{-1}$. Initially we set soil conductivity to zero and assume that permeability is~1.0. The reflection coefficients $r_v$ and $r_h$ for vertical and horizontal polarizations are then
\begin{subequations}
\begin{align}
    {r_v}&=  \frac{\epsilon_r {\rm cos}{\theta_i}-\sqrt{\epsilon_r-{\rm sin}^2\theta_i} }
{\epsilon_r{\rm{cos}}\theta_i+\sqrt{\epsilon_r -{\rm{sin}}^2\theta_i}} \thinspace,\\
    {r_h}&= \frac{{\rm{cos}}\theta_i-\sqrt{{\epsilon_r}-{\rm{sin}}^2\theta_i}}
    {{\rm{cos}}\theta_i+\sqrt{{\epsilon_r}-{\rm{sin}}^2\theta_i}} \thinspace,
\end{align}
\end{subequations}
where ${\theta}_i$ is the angle of incidence. $r_v$ and $r_h$ are ratios of reflected to incident field amplitudes. The reflectivities ${R_v}={r_v}^2$ and ${R_h}={r_h}^2$, which we will use later, are the ratios of reflected to incident powers.  For small values of conductivity, to good approximation, the brightness temperatures of the ground in the two polarizations are
\begin{subequations}
\begin{align}
    {T_v}&={(1-{{r_v}^2})T_p}\label{T_v} \thinspace,\\
    {T_h}&={(1-{{r_h}^2})T_p}\label{T_h} \thinspace,
\end{align}
\end{subequations}
where $T_p$ is the physical temperature of the ground, which we take to be 300~K. We plot these values in Figure~\ref{groundtemp} (together with other
quantities that will be discussed below).

There are several reasons to expect that the vertical and horizontal components of the ground emission are correlated. First, consider the emission process. A thermal signal is generated in lossy material below the ground surface as an unpolarized wave. This propagates towards the surface and it is the direction-dependent transmissivity/reflectivity of the ground/air interface that leads to the apparent polarization of the ground signal. The ground emission signal will be partly polarized, with polarized brightness $T_p$ and unpolarized brightness $T_u$, leading to fractional polarization
\begin{equation}\label{f_p}
{f_p}={\frac{T_p}{{T_p}+{T_u}}}={\frac{{T_v}-{T_h}}{{T_v}+{T_h}}} \thinspace.
\end{equation}

Equation \ref{f_p} shows that ${f_p}=0$ at vertical incidence and ${f_p}=1$ at grazing incidence. Ground emission from directly below the telescope is unpolarized, ground emission from the horizon is totally polarized, and at intermediate zenith angles is partially polarized. $T_v$ is always greater than $T_h$, so the ground emission is always partially vertically polarized. The apparent brightness temperature of the ground then has a dependence on angle of incidence that is intermediate between the two values suggested by equations \ref{T_v} and \ref{T_h}.

Furthermore, the ground surface is not smooth, and on rough ground the distinction between vertical and horizontal polarization is blurred, since the antenna sidelobe will intersect a rough ground surface at a range of incidence angles, not at a single angle. The effects are treated quantitatively by \cite{kerr90}. Using data from satellite remote sensing of the Earth's surface, these authors define a polarization coupling factor, $q$, as
\begin{equation}
{q}={0.35(1-{e^{{-0.6{\sigma_G}^2}{\nu}}})} \thinspace,
\label{coupling}
\end{equation}
where $\sigma_G$ is the rms deviation of the ground (cm) and $\nu$ is the frequency (GHz). The reflectivities, $R_{\rm rv}$ and $R_{\rm rh}$, of the rough ground for vertical and horizontal polarization are
\begin{subequations}
\begin{align}
    {R_{\rm rv}}&={q{R_h}+(1-q){R_v}{e}^{(-h{\rm{cos}}^2{\theta_i})}}\label{R_rv},\\
    {R_{\rm rh}}&={q{R_v}+(1-q){R_h}{e}^{(-h{\rm{cos}}^2{\theta_i})}}\label{R_rh}.
\end{align}
\end{subequations}
Here ${h}={(2k{\sigma_G})^2}$ is a roughness factor, where $k={2{\pi}/\lambda}$ is the wave number. Equation~\ref{coupling} and equations \ref{R_rv} and \ref{R_rh} indicate that surface roughness leads quickly to coupling between vertical and horizontal polarizations.  As the ground roughness increases, its reflectivity tends towards zero: it becomes a better absorber (and so a stronger emitter) and the effective ground temperature tends towards the physical ground temperature, about 300~K. These effects are illustrated in Figure~\ref{groundtemp}. This figure will apply to all frequencies for which our original assumption, ${{\epsilon}_r}=7$, remains valid; no significant change is expected across the band 1280 to 1750~MHz. A detailed discussion of Figure~\ref{groundtemp} can be found in Section~\ref{ground-obs}.

\begin{figure}
   \centerline{\includegraphics[width=0.48\textwidth]{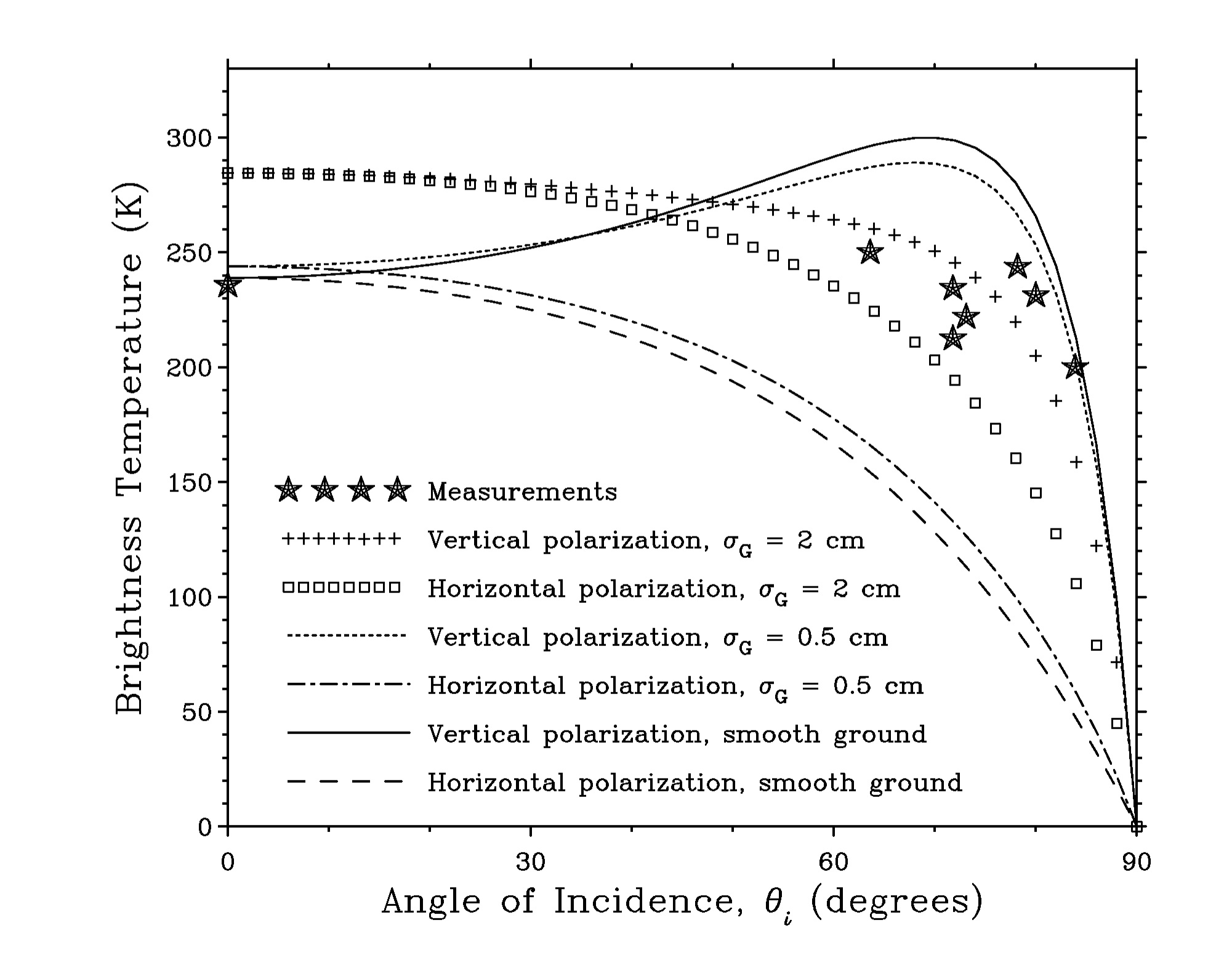}}
   \caption{Ground temperature as a function of angle of incidence in vertical
   and horizontal polarizations, shown for smooth ground, and rough ground with
   rms deviation 0.5~cm and 2~cm. The star symbols show measured points (see Section~\ref{ground-obs}).
   Note the measured point at ${\theta_i}={0^{\circ}}$. Values of ground
   temperature are valid for the entire frequency range 1250 to 1750~MHz.}
     \label{groundtemp}
\end{figure}

\subsection{Prediction of Spurious Polarization from Ground Emission}
\label{ground-predict}

We have computed the expected contribution of ground emission to $I$, $Q$, and $U$ based on the equations above. We adopted a greatly simplified profile for the ground, illustrated in Figure~\ref{groundprofile}, which approximates the topography around the telescope. We apply this radial profile to all azimuths, considering the ground to be symmetrical around the location of the telescope. The zone around the telescope to a viewing angle\footnote{The viewing angle is the angle, measured from the nadir, at which the emission enters the telescope.} of $69^{\circ}$ is flat with roughness ${\sigma_G}={0.5{\thinspace}{\rm{cm}}}$ (it consists of manicured lawns and paved surfaces). From ${\theta_i}={69^{\circ}}$ to ${\theta_i}={78^{\circ}}$ the ground is still flat, but rougher, with ${\sigma_G}={2{\thinspace}{\rm{cm}}}$. From ${\theta_i}={78^{\circ}}$ the ground begins to slope upwards at an angle of $10^{\circ}$ to approximate the hillsides, extending upwards at this angle to the mountaintops, at an elevation $10^{\circ}$ above the horizon. The roughness on the hillsides is also ${\sigma_G}={2{\thinspace}{\rm{cm}}}$.

\begin{figure}
   \centerline{\includegraphics[width=0.48\textwidth]{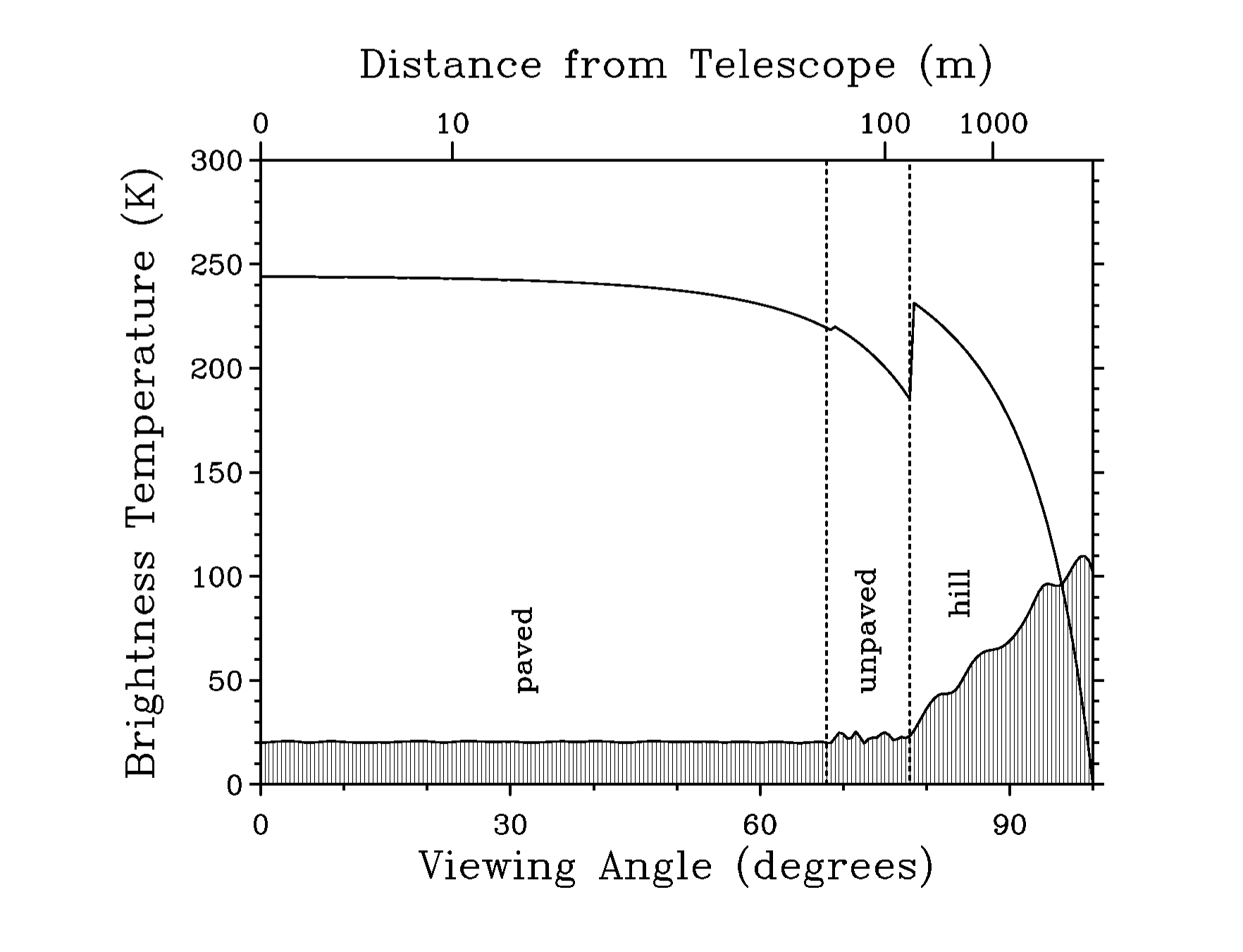}}
   \caption{The brightness temperature of the ground used in computing the telescope response to ground radiation.
   This is based on the hatched ground profile as shown at the bottom of the plot. The ground is flat to a viewing angle of $78^{\circ}$. The hill rises from there to a viewing angle of $100^{\circ}$. In the paved area ${{\sigma}_G}={0.5}~{\rm{cm}}$. In the unpaved area and on the hill
   ${{\sigma}_G}={2}~{\rm{cm}}$. The correspondence between viewing angle and distance from the telescope is shown by the labels at the top of the figure.}
     \label{groundprofile}
\end{figure}

We have calculated the signals expected in $I$, $Q$ and $U$ based on the ground model in Figure~\ref{groundprofile} together with the equations of Section~\ref{ground-pol}. At the same time we have calculated the signals expected if the ground is an unpolarized emitter with an effective temperature of 240~K with no variation with angle of incidence. The results are shown in Figure~\ref{ground1} together with measured values (see Section~\ref{ground-obs} for a description of the measurement method and for a justification of the use of 240~K as the effective temperature of the ground; see Section~\ref{compare} for detailed discussion). $I$, $Q$, and $U$, arising from ground emission, are shown as a function of zenith angle as the telescope moves along the meridian from zenith angle $-40^{\circ}$ (north of the zenith, corresponding approximately to declination $90^{\circ}$), to zenith angle $70^{\circ}$ (close to the southern limit of telescope operation at declination $-30^{\circ}$). This is the observing track used for the GMIMS observations.

\subsection{Ground Contribution Determined from Observations}
\label{ground-obs}

Also shown in Figure~\ref{groundtemp} are measured ground temperatures at 1.4~GHz from \citet{ande91}, made at 1420~MHz with an antenna with a beam of ${\sim}1.5^{\circ}$. Angles of incidence were derived from a topographical map. The measurement at ${\theta_i}={0^{\circ}}$ was made with a handheld horn. These measurements imply that the effective ground temperature is $\sim$240~K, somewhere between the temperatures expected for vertical and horizontal polarization (as predicted in Section~\ref{ground-pol}).

We interpret Figure~\ref{groundtemp} in the following way. The measurement at vertical incidence, ${\theta_i}={0^{\circ}}$, is believable. It was made with a handheld
horn antenna pointed vertically downward at the ground. It is credible that ${\sigma_G}\lessapprox{0.5\thinspace{\rm{cm}}}$ in the small patch of ground under the horn. However, this measurement is mostly irrelevant, because the spillover lobes do not "see" the ground directly below the telescope. Emission from this part of the ground does leak through the mesh into the receiver, but this accounts for only $\sim$10\% of ground emission. The spillover lobes mostly intersect more distant ground, far from the pavement and manicured lawns that surround the telescope, and the ground is rougher out there. ${\sigma_G}\approx{2\thinspace{\rm{cm}}}$ is quite a good approximation to ground in that zone, although the oblique angles of incidence need to be considered.  The rougher ground pushes the curves for vertical and horizontal polarization closer together. Furthermore, lines of sight never reach grazing incidence because the Observatory is in a bowl-shaped valley, and ${\theta_i}$ probably never exceeds ${80^{\circ}}$; the region relevant to spillover is ${30^{\circ}}\leq{\theta_i}\leq{80^{\circ}}$. The measurements lead us to believe that an average value of brightness temperature, the mean of vertical and horizontal, is quite appropriate, and the measured data support this view. Our estimate is consistent with the work of \citet{kalb10}, who quote a value for the albedo of the ground at 1420~MHz of 0.2, yielding a ground brightness temperature of $\sim$240~K.

\begin{figure}
\centerline{
\includegraphics[width=0.42\textwidth]
{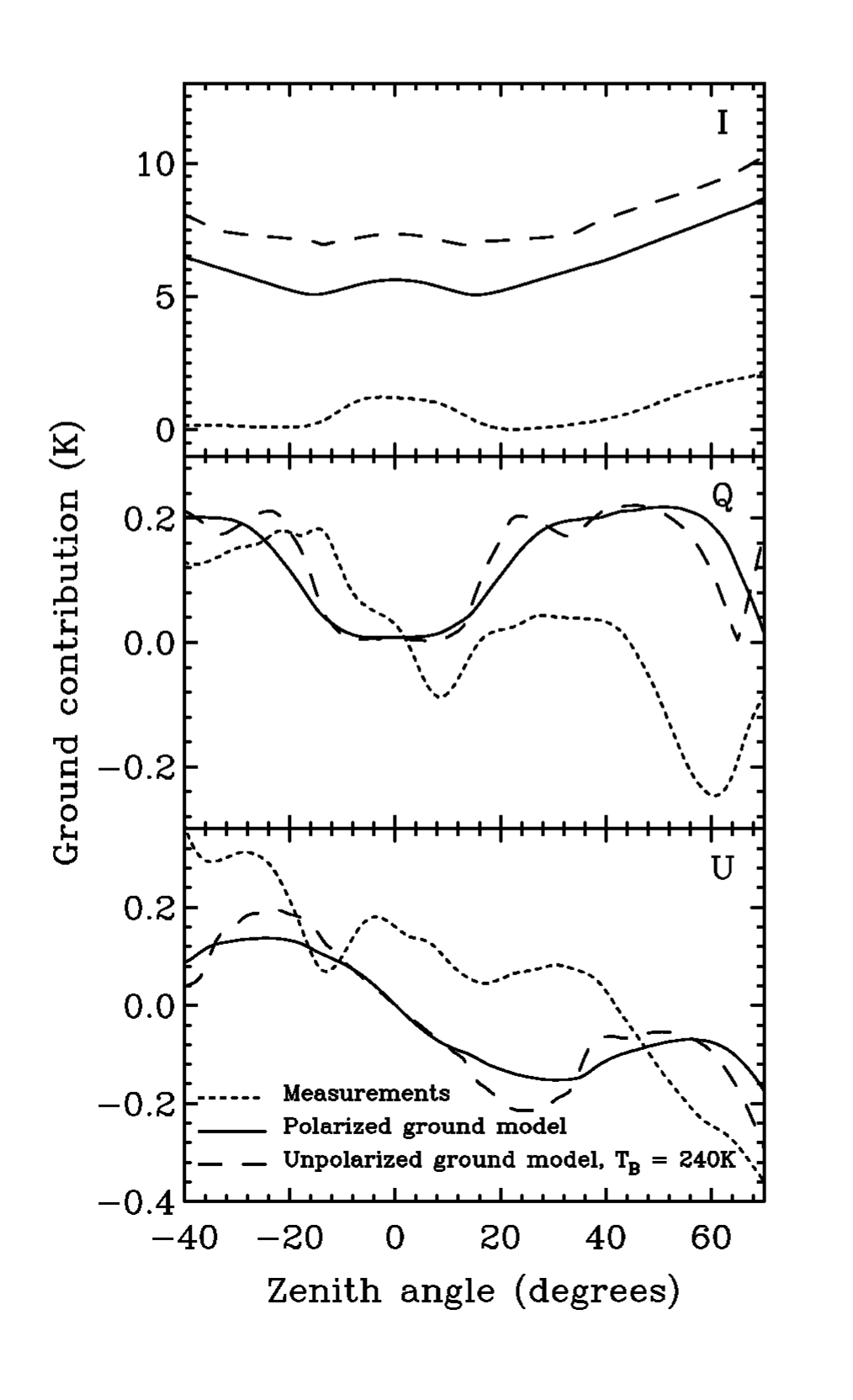}}
\caption{Computed contribution at 1400~MHz of ground radiation to Stokes
   parameters $I$, $Q$, and $U$ as a function of elevation. The solid lines show
   computed contributions if the ground emission is partially polarized, as
   discussed in Section~\ref{ground-pol}. The dashed lines show calculations
   assuming an unpolarized ground at an effective temperature of 240~K. The dotted
   curves are derived from measurements (see Section~\ref{ground-obs}).}
   \label{ground1}
\end{figure}

The GMIMS project, described in Section~\ref{intro}, provides us with measurements that we can compare with the results of our simulations. The observations were made using a meridian-nodding mode, where the Galt Telescope moved vertically at a rate of approximately $1^\circ$ per minute, covering the declination range $-30^{\circ}$ to $+87^{\circ}$.  Each such telescope track is referred to as a ``scan''. The equivalent zenith angle range for these ``scans'' is $79.5^{\circ}$ to the South, through zenith, to $40.5^{\circ}$ to the north. Earth rotation during a declination scan caused each observing path to be a diagonal line in the equatorial coordinate system. Successive tracks created individual diagonal lines until the entire sky was mapped with crossing tracks, with ${\sim}5{\times}10^5$ crossing points. After calibration against sources of known flux density the many crossing scans were reconciled using a ``basketweaving'' routine which iteratively deduced the best-fit zero level for each scan.

The output from this process was an image of the polarized signal from the sky plus the emission received from the ground and from the atmosphere. These contaminating signals were isolated by averaging across right ascension, for each declination, all signals recorded for the interval between right ascensions $04^{\rm h}$ and $08^{\rm h}$, an area of low sky brightness. It is safe to assume that the polarized signals are relatively low and the polarization angle will take almost all possible values, so the sky signal will average to zero over this large area, leaving only the contaminating signals. This process yielded curves of the spurious $Q$ and $U$ contributions produced by ground emission plus atmospheric emission as a function of elevation angle.

The total intensity (Stokes $I$) emission from the ground and atmosphere was estimated by stacking all signals in the same range of right ascension and taking the lower envelope of the measured signal. The known atmospheric contribution (see Section~\ref{atmos}) was subtracted to leave the ground contribution profile as a function of declination (elevation).

\subsection{Comparison of Calculated and Measured Ground Contributions}
\label{compare}

Figure~\ref{ground1} demonstrates that the spurious $Q$ and $U$ signals that arise from ground emission can reach magnitudes of 0.2 to 0.3~K. These values are significant, considering that the brightest polarized emission at 1.5~GHz is only $\sim$0.5~K. The spurious signals are lowest when the telescope is operating near the zenith; there is a high degree of symmetry in the radiation patterns, especially in $Q/I$ and $U/I$ (see Figure~\ref{rp}) and positive and negative responses fall on the ground in roughly equal measure. Examination of Figure~\ref{ground1} shows that there is relatively little difference between the contributions calculated on the basis of a partially polarized ground and on the basis of an unpolarized ground with an effective temperature of 240~K. The greatest effect of the dielectric properties of the ground is to reduce its effective temperature below its physical temperature, without significant change to the polarization properties of the ground radiation in the spurious $Q$ and $U$ signals. The measured ground contribution in $I$ is much lower than the calculated contribution. This is the result of the basketweaving process, which removes any uniform background level. This effect will have to be corrected in the GMIMS dataset, but is not a concern here.

The ground $I$ signal at the zenith is a local maximum, surrounded by minima $\sim$1~K lower at ${\pm}20^{\circ}$ on either side. This is easily understood in an intuitive way. At the zenith, a ring of spillover sidelobes lies uniformly on the ground. As the telescope is tipped, one side of the ring lifts off the ground while the contribution from the opposite side is relatively unchanged, and the ground contribution drops. Interestingly, this zenith `bump' was reproduced in the model of the unpolarized ground, but the bump did not appear with the polarized ground until the model of the polarized ground (Figure~\ref{groundprofile}) included a realistic characterization of the mountainsides. The most rapid changes in $Q$ and $U$ occur in the range of zenith angle between ${\sim}12^{\circ}$ and ${\sim}25^{\circ}$ (on both sides of the zenith). This is the result of interaction of the scatter cones with the emission from the surrounding mountains. The scatter cone maxima are at $68^{\circ}$ from the telescope boresight, and will intersect the top of the  mountains (at elevation $10^{\circ}$) at zenith angle ${\sim}12^{\circ}$.

\begin{figure*}
\centerline{
\includegraphics[width=0.72\textwidth,clip] {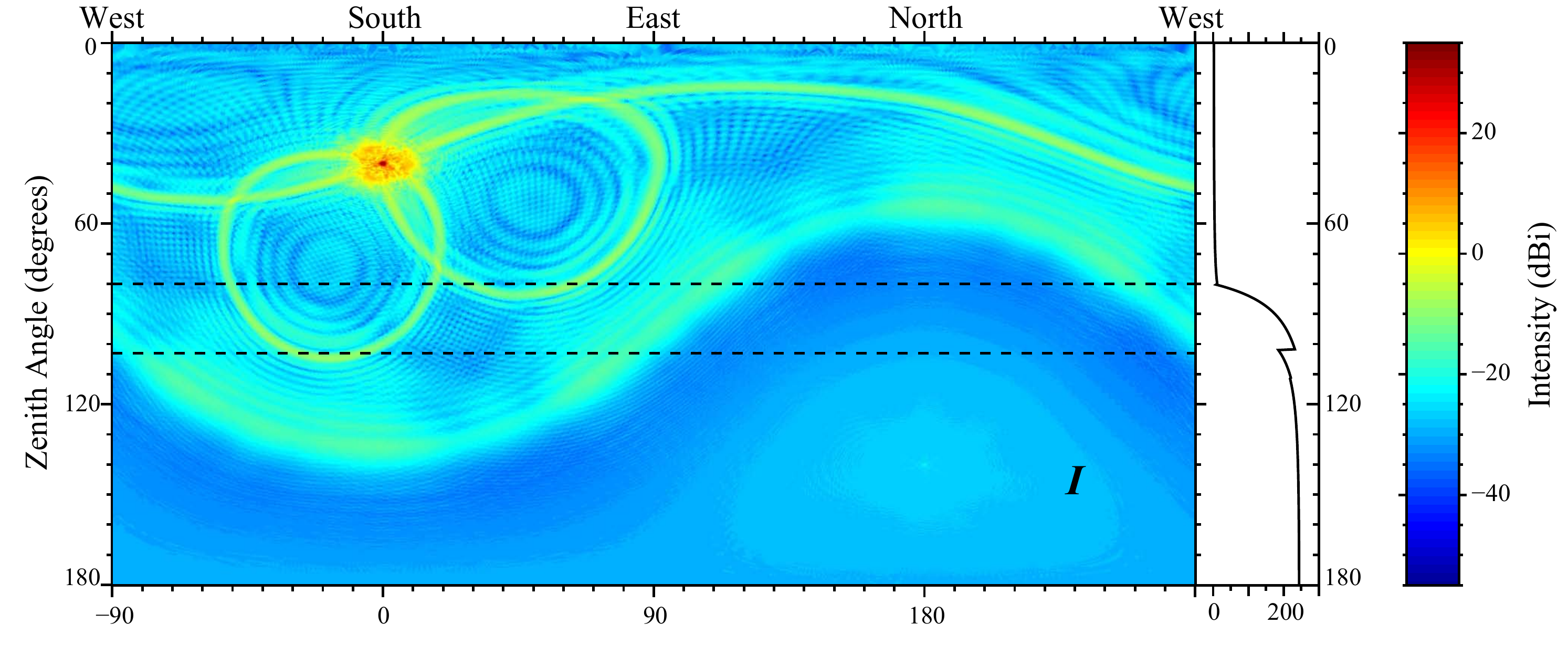}}
\centerline{
\includegraphics[width=0.72\textwidth,clip] {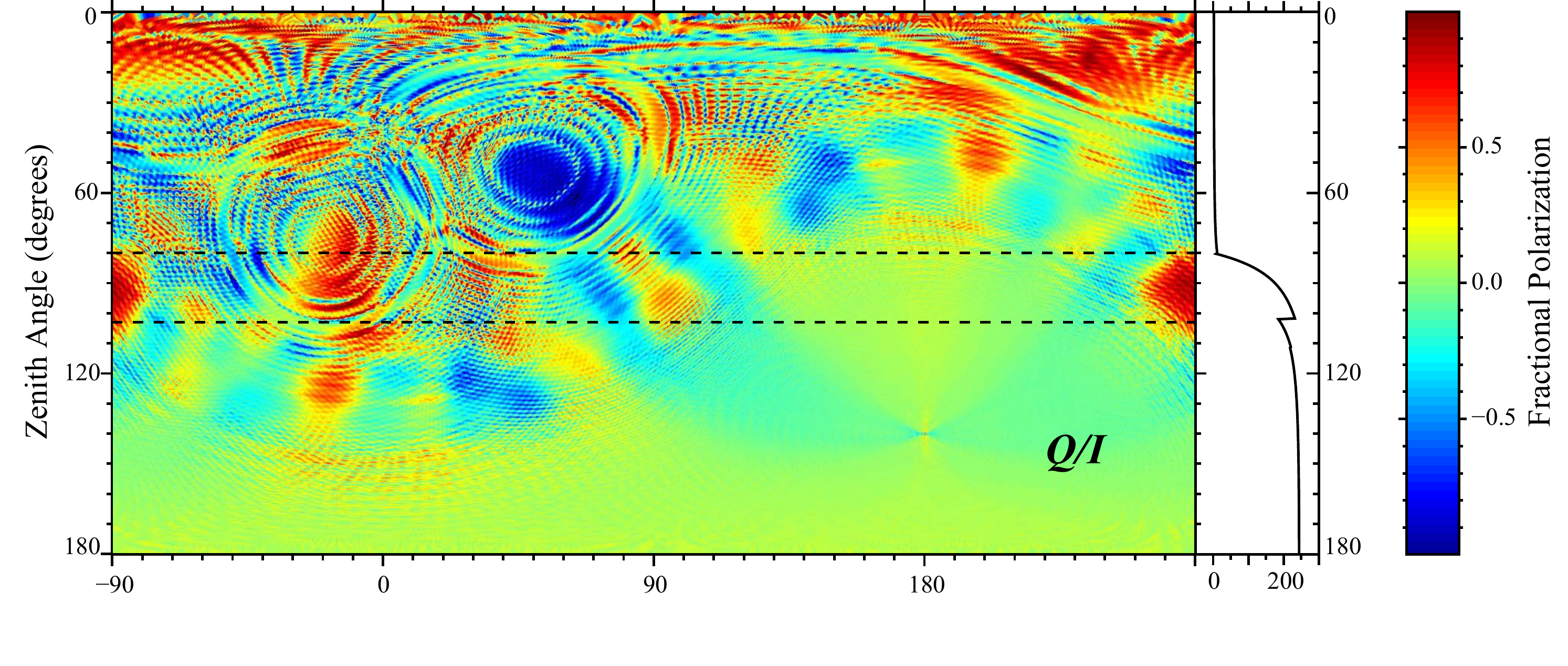}}
\centerline{
\includegraphics[width=0.72\textwidth,clip] {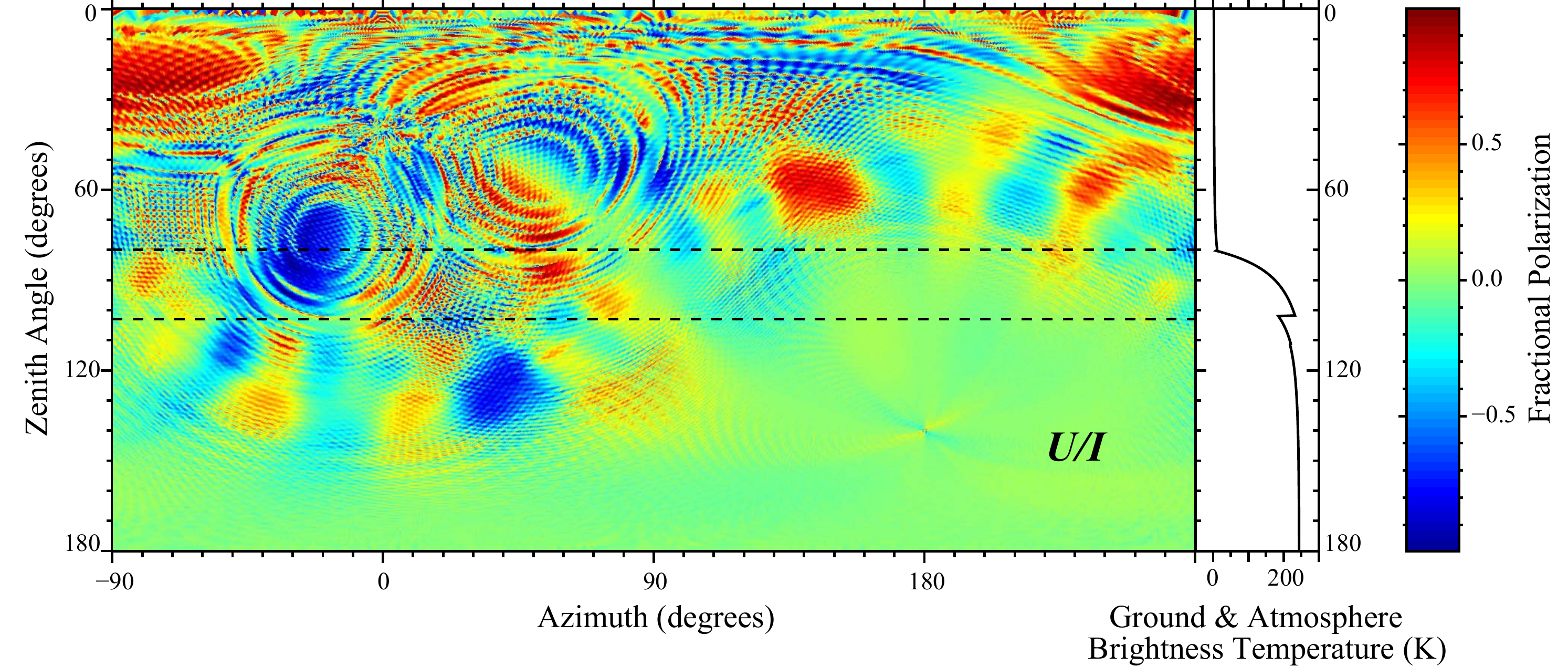}}
\caption{Illustrating the interaction of the antenna radiation pattern with ground and atmospheric emission. The left panels show $I$, $Q/I$, and $U/I$ in a horizontal coordinate system with the telescope directed at $40^{\circ}$ from the zenith (declination $9.5^{\circ}$) towards the south. The upper part of the radiation patterns sees the sky; the lower part sees the ground. The two dashed lines indicate, from upper to lower, the top and the bottom of the hill. The unit of $I$ is dB relative to isotropic and the color scale of the $Q/I$ and $U/I$ plots spans $-1$ to $+1$ as in Figure~\ref{rp}. The right panels show the brightness temperature of the surroundings as a function of zenith angle as in Figure~\ref{groundprofile}.
\label{ground2}}
\end{figure*}

As an illustration of the interaction of the radiation pattern with the surroundings of the antenna (ground and atmosphere) we have prepared Figure~\ref{ground2}. The radiation patterns (in $I$, $Q/I$, and $U/I$) are shown for the telescope directed at zenith angle $40^{\circ}$. This is the zenith angle where the $Q/I$ and $U/I$ signals are quite strong (Figure~\ref{ground1}). Looking at the $I$ pattern in Figure~\ref{ground2} we see that the spillover lobes will have a major influence on the ground contribution, simply because of their large solid angles. The spillover lobes in $Q/I$ and $U/I$ are broken up into alternating positive and negative responses, so the ground contributions tend to average out. The larger fraction of the ground contribution comes from the large lobes, uniform in sign, positive in the case of $Q/I$ and negative in $U/I$, that are associated with the scatter cones.

How successful have we been in reproducing the measured ground contributions to $Q$ and $U$? Figure~\ref{ground1} shows that we have the sense of the variation with zenith angle approximately right, but that there are still significant differences, especially in $Q$, and especially when pointing to the south (zenith angles $0^{\circ}$ to $80^{\circ}$). Neither the unpolarized ground model nor the polarized model comes close to the measurement.  Either our model of the ground is wrong, or our model of the strut that generates the scatter cone in this direction is wrong. This is the top strut (Figure~\ref{shadows}) the one that has no cables. Experiments showed that the ground model did affect the ground profiles in Figure~\ref{ground1}. A model based on the actual topography around the telescope might be useful, but there is a building close to the telescope, just north of it, and it is not clear what impact this has. Such an investigation is beyond the scope of this paper. If we want to rely on calculation of the spurious $Q$ and $U$ ground
contributions to make corrections to observations we would need to reach an accuracy of about $20\thinspace{\rm{mK}}$. We are at least a factor of 5 from achieving this goal.

\section{Atmospheric Emission}
\label{atmos}

Atmospheric emission contributes to system noise, especially at short wavelengths, entering the telescope through the main beam and the sidelobes. Emission from the atmosphere is unpolarized but can be converted into apparently polarized radiation by the instrumental polarization. We have calculated the contribution to polarized emission using the same routines as we applied to the problem of ground emission.

For this calculation the sky was represented as a thermal emitter whose temperature was dependent on direction. We used the equations given by \citet{gibb86} to calculate the atmospheric attenuation in dB/km. This was converted to a noise temperature using a scale height of 6 km and an average temperature of 260~K \citep{alln89}. The temperature at the zenith at 1.4 GHz is 2.0~K, and the variation with zenith angle, $z$, is taken as sec$\thinspace(z)$, the variation expected of a plane parallel slab of absorbing atmosphere.

The results are shown in Figure~\ref{atmo}. The variation of $Q$ and $U$ as a function of $z$ is similar in form to the variation of ground radiation, but inverted and much smaller in amplitude (the inversion can be seen clearly by comparing Figure~\ref{ground1} and Figure~\ref{atmo}). We can easily understand these effects. The effective temperature of the atmosphere is much lower than that of the ground, and the atmospheric contribution is correspondingly smaller (see below for a more quantitative discussion). The resemblance of form indicates that, just as the scatter cones dominate the appearance of the ground emission profile, so they dominate the atmospheric emission profile. The inversion of the shape between the ground and atmosphere profiles can be understood this way: as $z$ increases through $12^{\circ}$ the scatter cones lift off the ground, and must then encounter the brightest parts of the atmosphere; as the ground ontribution drops the atmospheric contribution rises.

At first sight the spurious polarization arising from atmospheric emission seems negligibly small, but it is worthwhile to go deeper into the data. If the ground temperature is 240~K and the atmospheric contribution is 2~K (the zenith value at 1.4~GHz) then we would expect the atmospheric contribution to be 1/120 of the ground contribution. The actual ratio of the two contributions is about 40 (from a comparison of the change in the $Q$ contribution at zenith angles ${\pm}20^{\circ}$ in Figure~\ref{ground1} and Figure~\ref{atmo}), which implies that atmospheric emission must be about 6~K, three times the zenith value. In fact sec$(z)$ reaches a value of 3 at ${z}{\approx}{70^{\circ}}$, so there is a ring of ``warmer'' atmosphere around the telescope at large zenith angles.  If the scatter cones are interacting with atmospheric emission this far from the zenith (just above the surrounding hills) then our simulation results can be understood. Spurious polarization arising from atmospheric emission is unlikely to be a significant factor at $\sim$1.5~GHz, but it could be a factor at higher frequencies where astronomical polarized signal levels are lower and atmospheric emission becomes more intense.

\begin{figure}
%\centerline{
\includegraphics[width=0.45\textwidth]
{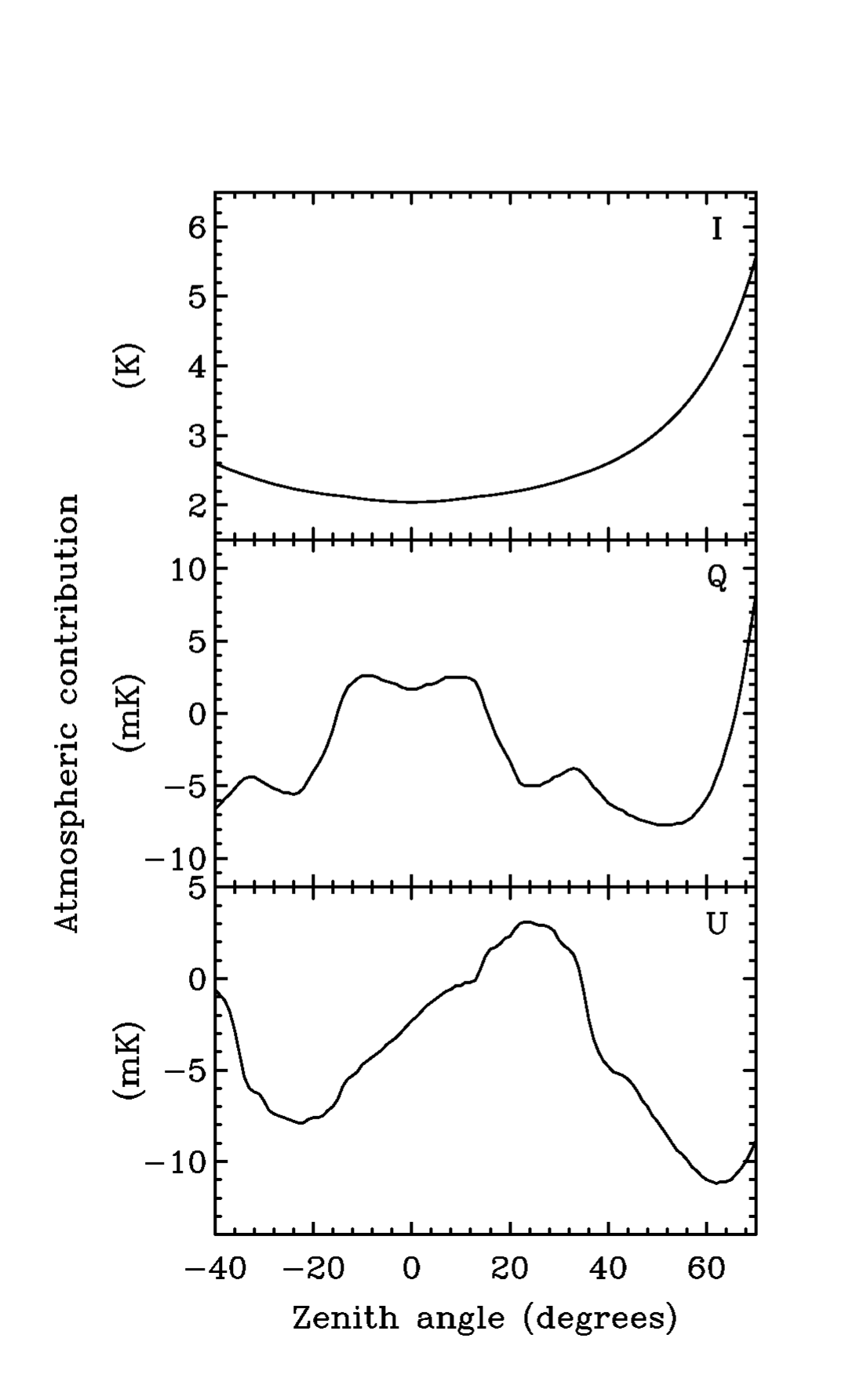}
\caption{Computed contribution at 1400~MHz of atmospheric emission for Stokes
   parameters $I$, $Q$, $U$, and $V$ as a function of elevation.}
   \label{atmo}
\end{figure}

\section{Concluding Discussion}
\label{disc}

We have used GRASP-10 to calculate the properties of a large radio telescope, and have compared the results of those calculations with measurements wherever possible. The GRASP-10 evaluation of aperture efficiency of the Galt Telescope has successfully predicted the frequency dependence of that parameter, but the measured values are approximately 6.5\% lower than the calculated value. This difference amounts to an error of 0.3~dB in the calculation of the gain, which is $\sim{50}$~dB. While this is good accuracy for an engineering tool, radio astronomy hopes to do better.

We also used simpler tools, ray tracing and geometrical optics, to examine aperture efficiency, and demonstrated that they too can provide useful accuracy and can provide useful insights into performance. We decomposed aperture efficiency into constituent efficiencies amenable to calculation on the basis of geometrical optics assumptions. The dominant effects in determining aperture efficiency are the illumination efficiency and blockage by the feed-support struts of the spherical wave from the feed: together they limit overall aperture efficiency to a maximum value of about 0.65.  The ray tracing code was of particular value in estimating spherical wave blockage because it was designed to handle the cigar-shape of the support legs. Spherical wave blockage is severe because the axes of the feed-support struts are only $34^{\circ}$ from the telescope axis, and a substantial part of the aperture is shadowed by the
struts.

The principal weakness of GRASP-10 in this application is its inability to model the struts precisely. GRASP-10 models the struts as straight metal cylinders, while in fact they are cigar-shaped fiber glass and two of them carry metal sheathed cables. We modelled the struts as metal cylinders of diameter 25.4~cm. If we had increased this by 3\%, a mere 8~mm, the gain would have decreased by about twice this, 6\%, bringing the calculation close to the measured result.

We have investigated the properties of a radio telescope over a very wide band, 1250 to 1750 MHz. The work that we describe here has established the absolute calibration of a wideband dataset. Previous work at these frequencies has been able to depend on absolutely calibrated horn measurements (e.g. the survey of \citealp{reic82} traced its calibration to \citealp{webs74}) but such measurements are confined to the narrow frequency bands allocated to radio astronomy. We have established an absolute scale of brightness temperature for the GMIMS survey. The overall error in this scale is 3\% but the relative error between any two frequencies in the band is less than this, about 1\%.

We have used GRASP-10 to investigate instrumental polarization in the far sidelobes. Signals from the ground (and to a lesser extent from the atmosphere) that are inherently unpolarized or slightly polarized can be converted to apparently strongly polarized signals. Our analysis has shown that the principal routes of entry of these spurious signals are the spillover lobes and the scatter cones that are generated by the feed-support struts. The most rapid changes in spurious polarization occur when the scatter cone sidelobes pass from the sky to the ground, or vice versa. For maximum stability of instrumental polarization, observations should be planned to avoid telescope pointings where the scatter cones come close to the horizon. Our predictions based on GRASP-10 agree in intensity with the measured results, but there are differences in detail. Again, the less than perfect representation in GRASP-10 of the feed-support struts is a problem. We also have the fact that our ground model is very simple; it does not take into account surrounding buildings, some of which are quite close to the telescope.

The ground is not a perfect absorber; it is a lossy dielectric. As a consequence its reflection and emission properties are polarization dependent. We have presented an analysis based on empirical data on ground polarization from studies using satellite-borne microwave radiometers. We conclude that the inherent polarization of ground emission is low at the angles that matter in the spillover lobes of the telescope. In fact our prediction of spurious polarization based on a polarized ground is not very different from our
prediction assuming an unpolarized ground, and in a comparison with the measured data it is hard to say which route gives the better prediction. We conclude that the best estimate of ground brightness temperature at $\sim$1.5~GHz is about 240~K independent of angle of incidence. \citet{kalb10} derive a very similar value at 1.42~GHz.

This conclusion does not conflict with observations that show that the Moon and planets are polarized at these frequencies (e.g. \citealp{zhan12,heil63}). A telescope observing the Moon or the planets samples only one angle of incidence using its main beam. In our case, however, the sidelobes of the Galt Telescope are sampling many different angles of incidence through many sidelobes. All these contributions are averaged and the ground appears to be almost unpolarized as seen by the sidelobes.

\acknowledgments

The Dominion Radio Astrophysical Observatory is a National Facility operated by the National Research Council Canada. The Global Magneto-Ionic Medium Survey is a Canadian project with international partners. The participation of KAD and MW in this research was supported by the Natural Sciences and Engineering Research Council (NSERC). Without the skilled work of Rob Messing and Ev Sheehan in maintaining the Galt Telescope this research would not have been possible. We thank Lloyd Higgs for his outstanding software, including the
ray-tracing program.

\bibliographystyle{aa}
\bibliography{polprop}

\end{document}